[1994/12/01]
\documentclass[screen]{aomart}
\usepackage[final]{graphicx}
\usepackage{amsfonts,bbm,amssymb,amsmath,slashed,mathrsfs}

\fancypagestyle{firstpage}{%
  \fancyhf{}%
  \if@aom@manuscript@mode
    \lhead{\begin{picture}(0,0)%
        \put(-21,-25){\usebox{\@aom@linecount}}%
      \end{picture}}
  \fi
  \chead{}%
    \cfoot{\footnotesize\thepage}}%
\doinumber{}

\chardef\bslash=`\\ 

\newtheorem[{}\it]{theorem}{Theorem}[section]
\newtheorem{corollary}[theorem]{Corollary}
\newtheorem{lemma}[theorem]{Lemma}

\theoremstyle{definition}
\newtheorem{definition}{Definition}[section]
\newtheorem{remark}{Remark}[section]
\newtheorem*[{}\it]{notation}{Notation}

\def\QEDopen{{\setlength{\fboxsep}{0pt}\setlength{\fboxrule}{0.2pt}\fbox{\rule[0pt]{0pt}{1.3ex}\rule[0pt]{1.3ex}{0pt}}}}
\def\QED{\QEDopen}

\def\Q.E.D{\hfill\QED}

\newcommand{\eval}[2][\right]{\relax
  \ifx#1\right\relax \left.\fi#2#1\rvert}

\title[$\mathcal{P}\neq\mathcal{NP}$]{Diagonalization of Polynomial-Time Deterministic Turing Machines via Nondeterministic Turing Machines}
\author[T. Lin]{Tianrong Lin}
\address{Hakka University\\ China}
\givenname{Tianrong}
\surname{Lin}
\copyrightyear{2023--}
\copyrightnote{}

\subject{primary}{matsc2020}{68Q15}
\subject{primary}{matsc2020}{68Q17}

\published{\formatdate{2008-04-02}}
\publishedonline{\formatdate{2007-11-12}}

\volumenumber{160}
\issuenumber{1}
\publicationyear{2008}
\papernumber{12}
\startpage{1}
\endpage{}

\version{2.1}

\mrnumber{MR1154181}
\zblnumber{0774.14039}
\arxivnumber{1234.567890}
\startpage{1}

\begin{document}

\begin{abstract}

The {\em diagonalization technique} was invented by Georg Cantor to show that there are more real numbers than algebraic numbers and is very crucial in {\em theoretical computer science}. In this work, we enumerate all of the polynomial-time deterministic Turing machines and diagonalize against all of them by a universal nondeterministic Turing machine. As a result, we obtain that there is a language $L_d$ not accepted by any polynomial-time deterministic Turing machines but accepted by a nondeterministic Turing machine running within time $O(n^k)$ for any $k\in\mathbb{N}_1$. Based on these, we further show that $L_d\in\mathcal{NP}$. That is, in this work, we present a proof that $\mathcal{P}$ and $\mathcal{NP}$ differ. Meanwhile, we show that there exists a language $L_s$ in $\mathcal{P}$, but the machine accepting it also runs within time $O(n^k)$ for all $k\in\mathbb{N}_1$. Lastly, we show that if $\mathcal{P}^O=\mathcal{NP}^O$ and on some rational base assumptions, then the set $P^O$ of all polynomial-time deterministic oracle Turing machines with oracle $O$ is not enumerable, thus demonstrating that the diagonalization technique ({\em via a universal nondeterministic oracle Turing machine}) will generally {\em not} apply to the relativized versions of the $\mathcal{P}$ versus $\mathcal{NP}$ problem.

\end{abstract}

\maketitle
\tableofcontents

\section{Introduction}
\label{sec:introduction}
  \vskip 0.3 cm

In $1936$, Turing's landmark $36$-page paper \cite{Tur37} opened the door to computer science, which has since evolved into numerous subfields such as {\em computability theory}, {\em formal language and automata theory}, {\em computational complexity theory}, {\em algorithm theory}, and so on. Turing's contributions, on the one hand, were so highly influential in the development of theoretical computer science that he is widely regarded as the father of theoretical computer science \cite{A4}. But on the other hand, although Turing's work initiated the study of theoretical computer science, he was not concerned with the efficiency of his machines, which is the main topic in computational complexity theory. In fact, Turing's concern \cite{Tur37} was whether they can simulate arbitrary algorithms given sufficient time (see e.g., \cite{Coo00}).

The {\em computational complexity theory} is a central subfield of the {\em theoretical foundations of computer science}, which mainly concerns the efficiency of Turing machines (algorithms) or the intrinsic complexity of computational tasks, i.e., focuses on classifying computational problems according to their resource usage and relating these classes to each other (see e.g., \cite{A5}). In other words, it specifically deals with fundamental questions such as what is feasible computation and what can and cannot be computed with a reasonable amount of computational resources in terms of time or space. In short, the theory formalizes this intuition to study these problems and quantify their computational complexity. 
   
There are different measures of difficulty to the computing, such as the amount of communication, the most basic and fundamental of which that appear particularly important are perhaps time and space. The fundamental measure of time opened the door to the study of the extremely expressive time complexity class $\mathcal{NP}$, one of the most important classical complexity classes, i.e., nondeterministic polynomial-time. This class comprises languages that can be computed in polynomial time by nondeterministic Turing machines. The famous Cook-Levin theorem \cite{Coo71,Lev73} shows that this class has complete problems, which states that the {\em Satisfiability} is $\mathcal{NP}$-complete, i.e., {\em Satisfiability} is in $\mathcal{NP}$ and any other language in $\mathcal{NP}$ can be reduced to it in polynomial time. This famous result also opened the door to research into the rich theory of $\mathcal{NP}$-completeness \cite{Kar72}. 

The famous $\mathcal{P}$ versus $\mathcal{NP}$ problem, which unquestionably has caught the interest of the mathematical community (see e.g., \cite{Coo03,BCSS98,Sma00}), is a major open question in the theoretical computer science community, specifically, in the computational complexity theory community. The question asks whether every problem in $\mathcal{NP}$ can also be solved in polynomial time by a deterministic Turing machine. It appeared explicitly for the first time in the papers of Cook \cite{Coo71}, Karp \cite{Kar72}, and Levin \cite{Lev73}. From the point of view in \cite{Bus12}, one reason why this problem is catching the interest of the mathematical community \cite{BCSS98,Sma00,Coo03} is that $\mathcal{P}=\mathcal{NP}$ could make the practice of mathematics too easy. Since mathematical research could be automated by formalizing mathematical questions completely and then blindly searching for proofs of conjectured mathematical statements. If $\mathcal{P}=\mathcal{NP}$, this process could succeed whenever proofs are not too large. This would obviously be a major change in the practice of mathematics! It is also worth focusing on that the relationships between the notion of {\em one-way functions} (OWFs) \cite{DH76} in {\em modern cryptography} (i.e., {\em one-way functions in the average-case model}) and the $\mathcal{P}$ versus $\mathcal{NP}$ problem are interesting topics and are discussed in a standard cryptographic textbook \cite{Gol01} by Goldreich. In addition, the relationships between {\em one-way functions in the worst-case model} and the $\mathcal{P}$ versus $\mathcal{NP}$ problem are also discussed in \cite{Pap94} (see p. 281--284 in \cite{Pap94}, or \cite{Ko85,GS88}). For more other details, the reader can find that the reference \cite{Wig07} also contains a detailed background about it.
   
There are several different and equivalent formulations of the $\mathcal{P}$ versus $\mathcal{NP}$ problem; the interested reader is referred to \cite{May04}. In particular, the logical characterizations of the question ``is $\mathcal{P}$ a proper subset of $\mathcal{NP}$" can be reformulated as ``is existential second-order logic able to describe languages (of finite linearly ordered structures with nontrivial signature) that first-order logic with least fixed point cannot?"; see e.g., \cite{A1,May04}.

\begin{figure}[htb]
\center{\includegraphics[width=12.5 cm]{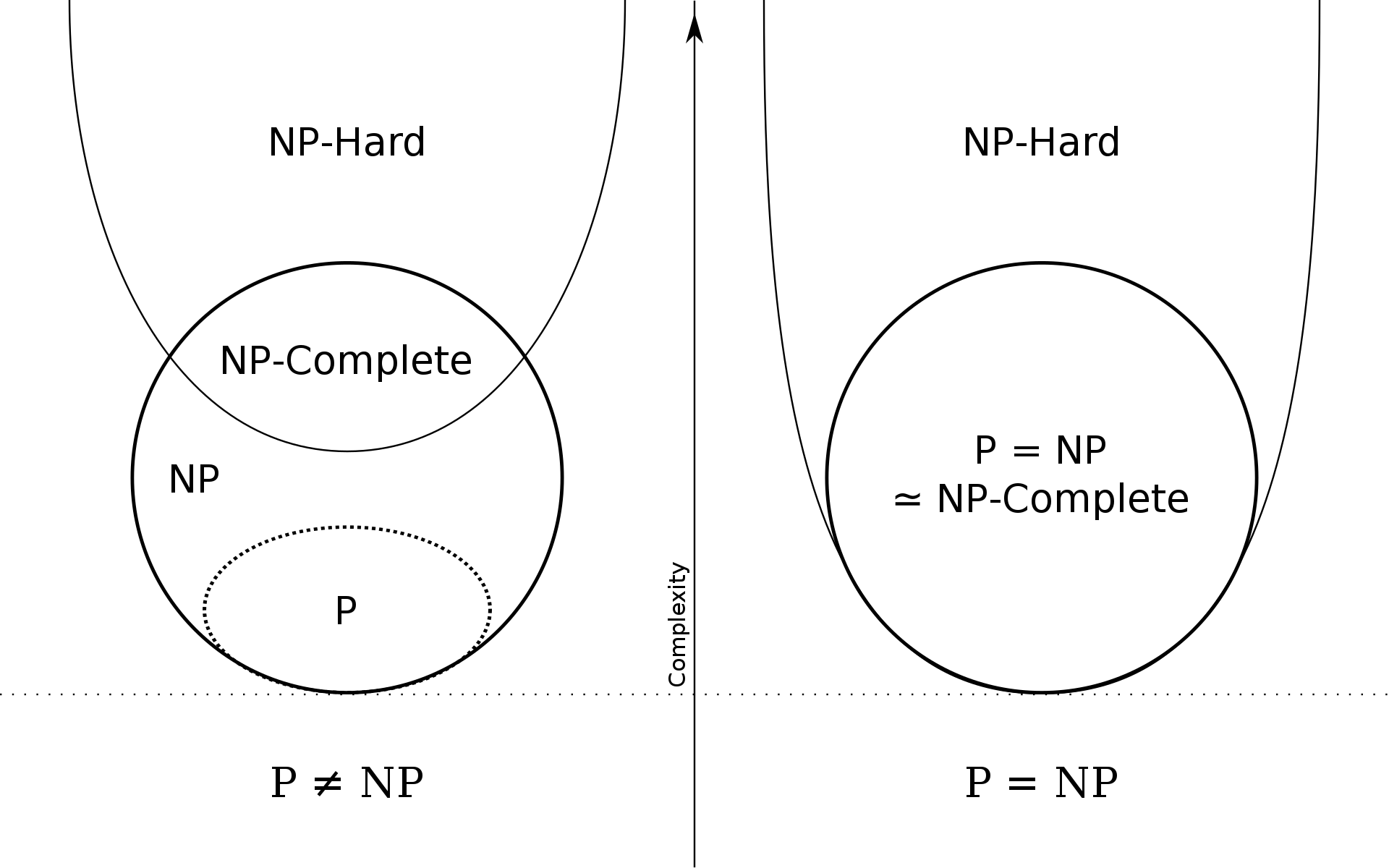}}
\caption{\label{1}Euler diagram for $\mathcal{P}$, $\mathcal{NP}$, $\mathcal{NP}$-complete, and $\mathcal{NP}$-hard sets of problems}
\end{figure}

So far, the exact relationship between the complexity classes $\mathcal{P}$ and $\mathcal{NP}$ is unknown. The figure \ref{1} below from \cite{A1} illustrates the Euler diagram (i.e., the relationship) for $\mathcal{P}$, $\mathcal{NP}$, $\mathcal{NP}$-complete, and $\mathcal{NP}$-hard sets of problems in two possibilities:

\vskip 0.3 cm
\vskip 0.3 cm

To the best of our knowledge, many of the methods used to attack the $\mathcal{P}$ versus $\mathcal{NP}$ problem (in the direction of $\mathcal{P}\neq\mathcal{NP}$) have been combinatorial or algebraic; for example, circuit lower bounds, diagonalization, and relativization \cite{BGS75}, but previously all of these attempts failed (see e.g., \cite{Coo00, Wig07, For09}). In 1997, Razborov and Rudich defined a general class of proof techniques for circuit complexity lower bounds (see \cite{Raz85}), called natural proofs \cite{RR97}. At the time, all previously known circuit lower bounds were natural, and circuit complexity was considered a very promising approach for resolving the $\mathcal{P}$ versus $\mathcal{NP}$ problem. However, in \cite{RR97}, Razborov and Rudich showed that if one-way functions exist, then no natural proof method can distinguish between $\mathcal{P}$ and $\mathcal{NP}$ (see e.g., \cite{A1}). Although one-way functions have never been formally proven to exist, most mathematicians believe that they do, and a proof of their existence would be a much stronger statement than $\mathcal{P}\neq\mathcal{NP}$. Thus it is unlikely that natural proofs alone can resolve the $\mathcal{P}$ versus $\mathcal{NP}$ problem (see e.g., \cite{A1}). But, maybe some day in the future, someone is going to present a proof of that $\mathcal{NP}\not\subseteq\mathcal{P}/poly$ by circuit lower bounds (for example, to prove {\em $3$-SAT} or other $\mathcal{NP}$-complete problems not in $\mathcal{P}/ploy$ by circuit lower bounds). In view of this, we believe that only God knows whether the circuit lower bounds really have a natural proofs barrier to the $\mathcal{P}$ versus $\mathcal{NP}$ problem.
   
More interestingly, the technique of diagonalization led to some early successes in computational complexity theory. In history, it was invented by Georg Cantor \cite{Can91} to show that there are more real numbers than algebraic numbers, then it was used by Alan Turing to show that there exists a sequence that is not computable (see \cite{Tur37}, p. 246), and then refined to prove computational complexity lower bounds by Hartmanis and Stearns \cite{HS65}. A typical theorem in this area is that more time/space buys more computational power \cite{HS65,Coo73,FS07,FS17,Pap94,Sip13,AB09}. For instance, there are functions computable in time $n^2$, say, which are not computable in time $n$. At present, we assume that the heart of such arguments is the existence of a universal Turing machine, which can simulate every other Turing machine with only a small loss in efficiency. \footnote{ At the end of the work, it will become clear that the notion of {\em Enumerable of Turing machines} is an important prerequisite for application of diagonalization techniques.}

As we mentioned earlier, the diagonalization technique was invented by Cantor \cite{Can91} to show that there are more real numbers than algebraic numbers, and early on it demonstrated powers in separating two complexity classes. There is an important doubt: can the diagonalization approach resolve the $\mathcal{P}$ versus $\mathcal{NP}$ problem? Researchers realized in the $1970$s that diagonalization alone may not be able to resolve the $\mathcal{P}$ versus $\mathcal{NP}$ problem; it needs other new techniques besides diagonalization (see e.g., \cite{Sip13, AB09}). But, at the end of the work, we will say that the above point of view is partially correct and leave aside why we think so.

It is interesting that the complexity class $\mathcal{NP}$ has a rich structure under the assumption that $\mathcal{P}$ and $\mathcal{NP}$ differ. In \cite{Lad75}, Lander constructed a language that is $\mathcal{NP}$-intermediate by the method of {\em lazy diagonalization} under the assumption that $\mathcal{P}\neq\mathcal{NP}$ (there are other problems that are suspected of being $\mathcal{NP}$-intermediate; for example, the graph isomorphism problem is believed to be at least not $\mathcal{NP}$-complete \cite{AK06}). We noted that the lazy diagonalization put the $\mathcal{NP}$-intermediate language accepted by no polynomial-time deterministic Turing machine, and our curiosity here, after reading \cite{Lad75} many times, is why Lander did not diagonalize against all polynomial-time deterministic Turing machines directly to see whether it is possible to give a solution of the $\mathcal{P}$ versus $\mathcal{NP}$ problem. Because it is a widespread belief that $\mathcal{P}$ and $\mathcal{NP}$ are different, we naturally would consider whether we can diagonalize against all of the polynomial-time deterministic Turing machines by a universal nondeterministic Turing machine to produce a language accepted by no polynomial-time deterministic Turing machine but accepted by some nondeterministic Turing machine. Once such a language that is not in $\mathcal{P}$ has been constructed by diagonalization techniques, we can do the second half of the work to see whether this language is in $\mathcal{NP}$ or not.

Before introducing the main results, let us stress that the definitions of almost all of the formal concepts involved in the subsection below will be postponed, and, due to our style of writing, we just suppose that the reader is familiar with these notions. To those who are not familiar with these concepts, we suggest proceeding to read the relevant definitions in Section \ref{sec:preliminaries} and in Section \ref{sec:relativization_barrier} first.
  
\vskip 0.3 cm
\subsection{Main Results}
\label{subsec:main_results}
\vskip 0.3 cm
    
We put our aforementioned curiosities, or ideas, into practice in this work. We first enumerate all polynomial-time deterministic Turing machines and then diagonalize against all of them by a universal nondeterministic Turing machine.\footnote{ It is much more suitable to say that we first present a way to enumerate all of the polynomial-time deterministic Turing machines.} Generally, complexity theorists used universal deterministic Turing machines to diagonalize against a list of deterministic Turing machines (see e.g., \cite{HS65,Pap94,AB09}), or applied universal nondeterministic Turing machines to diagonalize against a list of nondeterministic Turing machines; see e.g., \cite{AB09,For00,FS07,FS17,Zak83}. Using a universal nondeterministic Turing machine to diagonalize against a collection of deterministic Turing machines appears less often in the previous literature, except in the author's recent work \cite{Lin21}, and it is maybe a new attempt that may lead to success. As an amazing result, we obtain the following important theorem:

\begin{theorem}
\label{theorem1}
There exists a language $L_d$ that is not accepted by any polynomial-time deterministic Turing machines but is accepted by a nondeterministic Turing machine. Furthermore, this nondeterministic Turing machine runs within time $O(n^k)$ for any $k\in\mathbb{N}_1$. By this, it can be shown that $L_d\in \mathcal{NP}$.
\end{theorem}
   
 From which it immediately follows that:
\begin{corollary}
\label{corollary1}
$\mathcal{P}\neq \mathcal{NP}$.
\end{corollary}
   
\noindent and
\begin{corollary}
\label{corollary2}
$\mathcal{P}\neq \mathcal{PSPACE}$.
\end{corollary}
  
In order to contrast with the construction in the proof of Theorem \ref{theorem1}, we also design a universal deterministic Turing machine $M'_0$ that accepts a language in $\mathcal{P}$, and this machine also runs within time $O(n^k)$ for all $k\in\mathbb{N}_1$. In contrast, the deterministic universal Turing machine $M'_0$ does not perform any diagonalization operation. More interestingly, we conjecture that the language accepted by this deterministic universal Turing machine $M'_0$ cannot be accepted by any deterministic $O(n^i)$ time-bounded Turing machine for fixed $i\in\mathbb{N}_1$. Thus, the following theorem is an interesting by-product of our similar but different construction in proof of Theorem \ref{theorem1}:
   
\begin{theorem}
\label{theoremByProduct}
There exists a language $L_s$ accepted by a universal deterministic Turing machine $M'_0$ that is of time complexity $O(n^k)$ for all $k\in\mathbb{N}_1$. Further, $L_s$ is in $\mathcal{P}$.
\end{theorem}
   
Now, let us turn to the ``Relativization Barrier."
    
For any oracle $X$, we denote by $\mathcal{P}^X$ the class of languages recognized by polynomial-time deterministic oracle Turing machines with oracle $X$, and we denote by $\mathcal{NP}^X$ the class of languages accepted by polynomial-time nondeterministic oracle Turing machines with oracle $X$. More intuitively, let $P^X$ denote the set of all polynomial-time deterministic oracle Turing machines with oracle $X$, and $NP^X$ the set of all polynomial-time nondeterministic oracle Turing machines with oracle $X$, respectively.

In $1975$, Baker, Gill, and Solovay \cite{BGS75} presented a proof of that:\footnote{ The proof is via $\mathcal{PSPACE}=\mathcal{NPSPACE}$, whose proof is by the fact that the space is reusable but not valid for time, obviously.}
$$
\text{There is an oracle $A$ for which } \mathcal{P}^A=\mathcal{NP}^A.
$$
      
What are the implications of the above relativized result for the corresponding unrelativized question? A number of interesting ideas have been proposed, as observed by Ko (see \cite{Ko85}): for example, Baker, Gill, and Solovay \cite{BGS75} suggested that their result implies that ordinary diagonalization techniques are not capable of proving $\mathcal{P}\neq\mathcal{NP}$ (similar perspectives also followed in \cite{Sip13,BC94}). However, Kozen \cite{Koz78} disagreed with this point of view. Hartmanis and Hopcroft \cite{HH76} pointed out the possibility of the axiomatic independence of the question $\mathcal{P}\overset{?}{=}\mathcal{NP}$. Bennett and Gill \cite{BG81} showed that $\mathcal{P}^B=\mathcal{R}^B$ but $\mathcal{P}^B\neq\mathcal{NP}^B$ relative to a random oracle $B$. Based on their results, they proposed the random oracle hypothesis: for an ``acceptable" relativized statement $S^A$, $S^{\emptyset}$ is true if and only if $S^A$ is true with probability $1$ when $A$ is random.

To this day, almost all complexity theory experts notice that the proof techniques used to prove $\mathcal{P}\neq EXP$, i.e., {\em diagonalization techniques}, would also `apply verbatim' if we added an arbitrary oracle $O$. Thus, for any oracle $O$, we have $\mathcal{P}^O\neq EXP^O$. However, if we used similar techniques to show that $\mathcal{P}\neq\mathcal{NP}$, then it would also follow that $\mathcal{P}^O\neq \mathcal{NP}^O$ for all oracle $O$, which contradicts the result of \cite{BGS75} above. This is the so-called ``Relativization Barrier," which almost all complexity theory experts think that any proof technique leading to $\mathcal{P}\neq\mathcal{NP}$ should overcome. Simultaneously, this is also the most notable difficulty on the road to attacking this problem in the direction of $\mathcal{P}\neq\mathcal{NP}$ prior to our work. Based on the original perspectives given in \cite{BGS75}, it seems that the motivation of \cite{BGS75} is to prove that $\mathcal{P}\neq\mathcal{NP}$ implies that $\mathcal{P}^O\neq\mathcal{NP}^O$ for all oracle $O$, via relativization, i.e., a proof technique invariant to adding oracles, but nevertheless we prefer to regard that $\mathcal{P}\neq\mathcal{NP}$ is not necessarily a necessary and sufficient condition for $\mathcal{P}^O\neq\mathcal{NP}^O$ for all oracle $O$.\footnote{ For example, in the case where the set of $P^O$ is not enumerable, as will be explained below in detail.} Likewise, we prefer the point of view that the conclusions \cite{BGS75,Yao85} (i.e., there exists an oracle $A$ such that $\mathcal{P}^A\ne\mathcal{NP}^A$) imply that $\mathcal{P}^A=\mathcal{NP}^A$ for all oracle $A$ is not necessarily a necessary and sufficient condition for $\mathcal{P}=\mathcal{NP}$.

On the other hand, the role of relativization in complexity theory is interesting and important and has been a central theme in complexity theory for almost two decades. Following the work \cite{BGS75}, much effort was expended to find contradictory relativizations for other open problems in complexity theory. A major part of this effort revolved around obtaining relativized results about the polynomial time hierarchy and its relationship to other classes; see e.g., \cite{HCCRR93}. For example, contradictory relativizations for various unsolved problems regarding the polynomial hierarchy were found (see e.g., \cite{HCCRR93}). These and other results led to a strong belief that problems with contradictory relativization are very hard to solve and are not amenable to current proof techniques, i.e., the solutions of such problems are beyond the current state of mathematics (see e.g., \cite{Hop84, HCCRR93}).

In addition to the ``Relativization Barrier," some theorists in relativized worlds also developed the so-called ``Algebrization Barrier;" See \cite{AW09}. Also, the reader can find some interesting discussions about these barriers in the reference \cite{AB18}.

Now let us return to the proof techniques of {\it diagonalization} once again. Cantor's diagonal process, also called the diagonalization argument, was published in $1891$ by Georg Cantor \cite{Can91} as a mathematical proof that there are infinite sets that cannot be put into one-to-one correspondence with the infinite set of positive numbers, i.e., $\mathbb{N}_1$ defined in the following Section \ref{sec:preliminaries}. The technique of diagonalization was first used in computability theory in the $1930$s by Turing \cite{Tur37} to show that there exists undecidable language. In {\em computational complexity theory}, in their seminal paper \cite{HS65}, Hartmanis and Stearns employed the diagonalization proof to give time hierarchy. For more other summarization about it, please consult \cite{For00} for a survey. On the other hand, Arora and Barak \cite{AB09} (see e.g., page $73$ in \cite{AB09}, which is a textbook in {\em computational complexity theory}) regard that ``{\em diagonalization}" is any technique that relies solely upon the following properties of Turing machines:\\
\vskip 0.15cm
\indent {\bf I:} The existence of an effective representation of Turing machines by strings.\\
\indent {\bf II:}  The ability of one Turing machine simulate any other without much overhead in running time or space.
\vskip 0.3cm

Thus, Arora and Barak, the authors of the computational complexity textbook \cite{AB09}, and other complexity theorists think of these properties as also applicable for oracle Turing machines and further regard that to show $\mathcal{P}\neq\mathcal{NP}$ requires some other properties in addition to the properties {\bf I} and {\bf II} as stated above. However, we would like to say that this kind of perspective is not absolutely (fully) correct either. Notice also that a similar point of view was followed in \cite{Sip13,BC94}, i.e., the references \cite{Sip13} and \cite{BC94} hold the same viewpoint that if we could prove that $\mathcal{P}$ and $\mathcal{NP}$ were different by diagonalizing, then we could conclude that they are different relative to any oracle as well. Interestingly enough, we will demonstrate the key points besides the aforementioned assumptions {\bf I} and {\bf II} when diagonalization techniques are applicable such that we can overcome the so-called ``Relativization Barrier."

But at the moment, for the convenience of the reader, let us first quote a mathematical definition of an enumeration of a set, which appears in modern mathematics textbooks such as \cite{Rud76}, as follows:

\begin{definition}[\cite{Rud76}, p. 27, Definition 2.7]\footnote{ In Cantor's terminology, the enumeration of something is the ``sequence" of something. We should be clear that only enumerable sets have enumerations. And by the term ``{\em enumerable}", Turing refers to \cite{Hob21}, p. 78. That is, {\em an aggregate (i.e., set) that contains an indefinitely great number of elements is said to be enumerable, or countable, when the aggregate is such that a $(1,1)$ correspondence can be established between the elements and the set of integral numbers $$1,2,3,\cdots,$$} i.e., $\mathbb{N}_1$. We can simply deem that an enumeration of an enumerable set $T$ is just a function $$e:\mathbb{N}_1\rightarrow T$$ that is surjective; or equivalently, it is an injective function $$e':T\rightarrow \mathbb{N}_1,$$ meaning that every element in $T$ corresponds to a different element in $\mathbb{N}_1$. See \cite{Tur37}, Section of {\em Enumeration of computable sequences}.}
\label{definition1}
By an enumeration of set $T$, we mean a function $e$ defined on the set $\mathbb{N}_1$ of all positive integers. If $e(n)=x_n\in T$, for $n\in\mathbb{N}_1$, it is customary to denote the enumeration $e$ by the symbol $\{x_n\}$, or sometimes by $x_1$, $x_2$, $x_3$, $\cdots$. The values of $e$, that is, the elements $x_n\in T$, are called the {\em terms} of the enumeration.
\end{definition}

So far, we have not explained what a set $P^O$ of deterministic oracle Turing machines with oracle $O$ {\em being not enumerable} means. But, with the assistance of the above Definition \ref{definition1}, a simple semantics for a set $P^O$ of deterministic oracle Turing machines with oracle $O$ {\em being not enumerable} can simply be explained as follows: there exists no enumeration of the set $P^O$. Or more precisely, there exists no function $e$ from the set of positive integers $\mathbb{N}_1$ to the set of $P^O$ that is surjective. Namely, there exists no function $e$ from $\mathbb{N}_1$ to $P^O$ such that for any deterministic oracle Turing machine $M^O$ in $P^O$, there is an element $i$ in $\mathbb{N}_1$ satisfying that $$e(i)=M^O. $$ 
        
{\em To convince the experts} who argue that the ``Relativization Barrier" is a real barrier that should be overcome when proving $\mathcal{P}\neq\mathcal{NP}$, we show the following important theorem that concerns oracle Turing machines, which is on purpose to demonstrate that the ``Relativization Barrier" is not really a barrier. To do so, of course, we should first suppose without loss of generality that polynomial-time deterministic (nondeterministic) oracle Turing machines can be effectively represented as strings (i.e., the above property {\bf I}), and further, there are universal nondeterministic oracle Turing machines that can simulate and flip the answers of other deterministic oracle Turing machines without much overhead in running time or space (i.e., the above property {\bf II}). Indeed, the next theorem being introduced in the following is our conclusion in this regard. 
    
Now, it is time for us to state our next theorem as follows:

\begin{theorem}\footnote{
The proof of this theorem, in fact, is similar to the proof of Cantor's theorem: there are infinite sets that can not be put into one-to-one correspondence with the set of positive integers, i.e., $\mathbb{N}_1$; see e.g., \cite{Gra94}. Furthermore, the argument of this theorem lies in the assumptions that (I) polynomial-time deterministic (nondeterministic) oracle Turing machines can be effectively represented as strings; (II) a universal nondeterministic oracle Turing machine exists that can simulate and flip the answers of other deterministic oracle Turing machines; and (III) the simulation of a universal nondeterministic oracle Turing machine to any deterministic oracle Turing machine can be done within $$O(T(n)\log T(n))$$ steps, where $T(n)$ is the time complexity of the simulated deterministic oracle Turing machine.\label{footnote6}
} 
\label{theorem2}
Let $P^O$ be the set of all polynomial-time deterministic oracle Turing machines with oracle $O$. Under some rational assumptions (i.e., the conditions {\bf I} and {\bf II} given in Subsection \ref{subsec:main_results}), and if $$\mathcal{P}^O=\mathcal{NP}^O, $$then the set $P^O$ is not enumerable. That is, the cardinality of $P^O$ is larger than that of $\mathbb{N}_1$ (card $P^O>$card $\mathbb{N}_1$).
\end{theorem}

It follows immediately from the above Theorem \ref{theorem2} that

\begin{corollary}\footnote{ In Turing's way, he first assumes that the computable sequences are enumerable, then applies the diagonal process. See \cite{Tur37}, Section $8$ of {\em Application of the diagonal process}.}
\label{corollary3}
If a set $T$ (of oracle Turing machines) is enumerable, then the {\em diagonalization} technique may be applicable. In other words, that $T$ (of oracle Turing machines) is enumerable is an important prerequisite for the application of {\em diagonalization techniques}.
\end{corollary}

\vskip 0.3 cm
\subsection{Our Approach}
\label{sec:our_approach}
\vskip 0.3 cm

The Cook-Levin Theorem is a well-known theorem stating that {\em Satisfiability} (SAT) is complete for the complexity class $\mathcal{NP}$ under polynomial-time many-one reductions. Starting from this, on the one hand, if one wants to prove that $\mathcal{P}$ and $\mathcal{NP}$ are identical, then she/he can try to design polynomial-time algorithms for SAT. On the other hand, if one wants to show that $\mathcal{P}$ and $\mathcal{NP}$ differ, then she/he may try to prove a super-polynomial lower bound for {\em Satisfiability}.

As a novel idea and attempt, we enumerate all polynomial-time deterministic Turing machines and then diagonalize against all of them with a universal nondeterministic Turing machine, thus obtaining a language $L_d$ not in $\mathcal{P}$. Based on the above work, we then carefully do an analysis for the language $L_d$, showing that this language is in $\mathcal{NP}$. To stress that it is ``novel" is because, in general, a common practice is to use a universal deterministic Turing machine to diagonalize against a list of deterministic Turing machines (see e.g., \cite{Pap94, AB09}), or to employ a universal nondeterministic Turing machine to diagonalize against a list of nondeterministic Turing machines; see e.g., \cite{AB09, For00, FS07,FS17, Zak83}. That using a universal nondeterministic Turing machine to diagonalize against all of the polynomial-time deterministic Turing machines is less common and can be seen as a new attempt. We would like to emphasize that this approach has seldom been tried before in the literature, except that recently we noted that in \cite{Sip92}, Sipser mentioned an idea (but this simple argument fails; see \cite{Sip92}, p. 605) to give a nondeterministic polynomial-time Turing machine that has an opportunity to run each of the deterministic polynomial-time Turing machines and arrange to accept a differing language.
    
We should also point out that our idea of diagonalization against all of the polynomial-time deterministic Turing machines by using a universal nondeterministic Turing machine was inspired by our recent work \cite{Lin21}. Objectively, the work \cite{Lin21} is the source of our in-depth understanding and application of diagonalization techniques in the domain of {\em computational complexity}, due to the fact that the idea of diagonalization against deterministic Turing machines with a nondeterministic Turing machine goes back to \cite{Lin21}, in which we use a universal nondeterministic $n$ space-bounded Turing machine to diagonalize against a collection of deterministic $n$ space-bounded Turing machines.
    
As we mentioned earlier, we do not know whether the circuit lower bounds \cite{Raz85} really have a natural proofs barrier to the $\mathcal{P}$ versus $\mathcal{NP}$ problem. Anyway, even if someday there were proof of that $\mathcal{NP}\not\subseteq\mathcal{P}/poly$ by circuit lower bounds, we still believe that the approach presented in this work is the simplest, since we do not like complicated proofs either, nor do we like to make things much more complicated.

\vskip 0.3 cm
\subsection{Related Work}
\label{sec:related_work}
\vskip 0.3 cm
  
As is well known, a central open question in {\em computational complexity theory} is the $\mathcal{P}$ versus $\mathcal{NP}$ problem, which is to determine whether every language accepted by some nondeterministic Turing machine in polynomial time is also accepted by some deterministic Turing machine in polynomial time. In this subsection, we will review its history and related works. With regard to the importance of the problem, we refer the reader to the references \cite{Coo00, Coo03, Wig07}.

In 1971, Cook \cite{Coo71} introduced a notion of $\mathcal{NP}$-completeness as a polynomial-time analog of c.e.-completeness, except that the reduction used was a polynomial-time analog of Turing reducibility rather than of many-one reducibility (see Chapter $7$ in \cite{Rog67} for {\em Turing reducibility}). Besides the first well-known $\mathcal{NP}$-complete problem of Satisfiability, Cook also showed in \cite{Coo71} that several natural problems, including $3$-SAT and subgraph isomorphism, are $\mathcal{NP}$-complete.

A year later, stimulated by the work of Cook \cite{Coo71} (according to the viewpoints in \cite{Kar72}), Karp \cite{Kar72} used these completeness results to show the celebrated conclusions that $20$ other natural problems are $\mathcal{NP}$-complete, forcefully demonstrating the importance of the subject. Thus far, there are many problems shown to be $\mathcal{NP}$-complete. See the excellent reference \cite{GJ79} on this subject. In his paper \cite{Kar72}, Karp also introduced the now standard notation $\mathcal{P}$ and $\mathcal{NP}$ and redefined $\mathcal{NP}$-completeness by using the polynomial-time analog of many-one reducibility, which has become standard. Meanwhile, Levin \cite{Lev73}, independently of Cook \cite{Coo71} and Karp \cite{Kar72}, defined the notion of ``universal search problem," similar to the $\mathcal{NP}$-complete problem, and gave six examples, which include {\em Satisfiability}.

Although the precise statement of the $\mathcal{P}$ versus $\mathcal{NP}$ question was formally defined in the $1970$s in his seminal paper \cite{Coo71} by Cook, there were previous inklings of the problems involved (see \cite{A1}). A mention of the underlying problem occurred in a $1956$ letter written by K.  G\"{o}del to J.  von  Neumann. G\"{o}del asked whether theorem-proving could be solved in quadratic or linear time (see e.g., \cite{Har89}). It is worth paying attention that, besides the classical version of the question, there is a version expressed in terms of the field of complex numbers, which has attracted the interest of the mathematics community \cite{BCSS98}.
    
For more details about the history of the $\mathcal{P}$ versus $\mathcal{NP}$ problem, we refer the reader to \cite{Sip92}, in which it provides a very detailed description. Also, for popular introductions, we refer the reader to the easy-to-understand book \cite{For13}, which provides a non-technical introduction to the $\mathcal{P}$ versus $\mathcal{NP}$ problem.
    
\vskip 0.3 cm
\subsection{Organization}
\label{sec:overview}
\vskip 0.3 cm
The rest of the work is organized as follows: For the convenience of the reader, in the next section we will review some notions closely associated with our discussions and fix some notation we will use in the following context. Also, some useful technical lemmas are presented. In Section \ref{sec:enumeration}, we provide a method to encode a polynomial-time deterministic Turing machine to an integer so that we can prove that the set of all polynomial-time deterministic Turing machines is enumerable. Section \ref{sec:diagonalization1} contains the construction of our nondeterministic Turing machine, which accepts a language $L_d$ not in $\mathcal{P}$. And Section \ref{sec:proof_ldinnp} overcomes the obstacle to prove the language accepted by our simulating machine is in $\mathcal{NP}$. Section \ref{sec:dontsuspect} is devoted to proving Theorem \ref{theoremByProduct}, which states that there is a language $L_s\in\mathcal{P}$ and the machine accepting it also runs within time $O(n^k)$ for all $k\in\mathbb{N}_1$. In Section \ref{sec:relativization_barrier}, we focus our attention on overcoming the so-called ``Relativization Barrier." Finally, we draw some conclusions in the last section.
    
\vskip 0.3 cm
\section{Preliminaries}
\label{sec:preliminaries}
\vskip 0.3 cm

In this section, we describe the notation and notions needed in the following context. We would like to point out that our style of writing\footnote{ But merely the style of writing. Indeed, if we describe the preliminaries in the author's own words, it will appear rather verbose.} in this section is heavily influenced by that in Aho, Hopcroft, and Ullman's book \cite{AHU74}.

Let $\mathbb{N}$ denote the set of natural numbers $$\{0,1,2,3,\cdots\}$$ where $+\infty\not\in\mathbb{N}$. Furthermore, $\mathbb{N}_1$ denotes the set of $$\mathbb{N}-\{0\},$$ i.e., the positive integers. It is clear that there is a bijection between $\mathbb{N}$ and $\mathbb{N}_1$. To see this, just let the bijection be $$n\mapsto n+1,$$ where $n\in\mathbb{N}$ and $n+1\in\mathbb{N}_1$.

The big $O$ notation indicates the order of growth of some quantity as a function of $n$ or the limiting behavior of a function. For example, that $S(n)$ is big $O$ of $f(n)$, i.e.,
$$
S(n)=O(f(n)),
$$ 
means that there exist a positive integer $N_0$ and a positive constant $M$ such that
$$
S(n)\leq M\times f(n)
$$
for all $n>N_0$. 

The big $\Omega$ notation also indicates the limiting behavior of a function of $n$ with different means. Specifically, that $t(n)$ is big $\Omega$ of $g(n)$, i.e., $$t(n)\in\Omega(g(n)),$$ means that there exists a positive integer $N_0$ and a positive constant $c$ such that $$t(n)>c\times g(n)$$ for all $n>N_0$.

The computation model we use here is the {\em Turing machines} as it is defined in standard textbooks such as \cite{HU69,HMU06}. Here, we follow the standard definition presented in \cite{AHU74}:

\begin{definition}[$k$-tape deterministic Turing machine, p. $26$, $27$ in \cite{AHU74}]
\label{definition2}
A $k$-tape deterministic Turing machine (shortly, DTM) $M$ is a seven-tuple $(Q,T,I,\delta,\mathbbm{b},q_0,q_f)$ where:
\begin{enumerate}
\item {$Q$ is the set of states.}
\item {$T$ is the set of tape symbols.}
\item {$I$ is the set of input symbols; $I\subseteq T$.}
\item {$\mathbbm{b}\in T-I$ is the blank.}
\item {$q_0$ is the initial state.}
\item {$q_f$ is the final (or accepting) state.}
\item {$\delta$ is the next-move function, which maps a subset of $Q\times T^k$ to $$Q\times(T\times\{L,R,S\})^k.$$
That is, for some $(k+1)$-tuples consisting of a state and $k$ tape symbols, it gives a new state and $k$ pairs, each pair consisting of a new tape symbol and a direction for the tape head. Suppose $$\delta(q,a_1,a_2,\cdots,a_k)=(q',(a'_1,d_1),(a'_2,d_2),\cdots,(a'_k,d_k)),$$ and the deterministic Turing machine is in state $q$ with the $i$th tape head scanning tape symbol $a_i$ for $1\leq i\leq k$. Then in one move the deterministic Turing machine enters state $q'$, changes symbol $a_i$ to $a'_i$, and moves the $i$th tape head in the direction $d_i$ for $1\leq i\leq k$.}
\end{enumerate}
\end{definition}

\vskip 0.3 cm

The notion of a nondeterministic Turing machine is similar to that of a deterministic Turing machine, except that the next-move function $\delta$ is a mapping from $Q\times T^k$ to subsets of $Q\times(T\times\{L,R,S\})^k$, stated as follows:

\begin{definition}[$k$-tape nondeterministic Turing machine,  p. $365$  in \cite{AHU74}]
\label{definition3}
A $k$-tape nondeterministic Turing machine (shortly, NTM) $M$ is a seven-tuple $(Q,T,I,\delta,\mathbbm{b},q_0,q_f)$ where all components have the same meaning as for the ordinary deterministic Turing machine, except that here the next-move function $\delta$ is a mapping from $Q\times T^k$ to subsets of $Q\times(T\times\{L,R,S\})^k$.
\end{definition}

In the following, we will refer to the Turing machine as both the deterministic Turing machine and the nondeterministic Turing machine. And we will sometimes use DTM (respectively, NTM) to denote a deterministic (respectively, nondeterministic) Turing machine.

Let $M(w)$ denote that Turing machine $M$ is on input $w$. If for every input $w$ of length $n$,\footnote{In the following context, we will use $|w|$ to denote the length of $w$.} all computations of $M$ end in less than or equal to $T(n)$ steps, then $M$ is said to be a deterministic (respectively, nondeterministic) $T(n)$ {\em time-bounded Turing machine}, or is said to be {\em of time complexity $T(n)$.}

The family of languages of deterministic time complexity $T(n)$ is denoted by ${\rm DTIME}[T(n)]$; the family of languages of nondeterministic time complexity $T(n)$ is denoted by ${\rm NTIME}[T(n)]$. The notation $\mathcal{P}$ and $\mathcal{NP}$ is defined respectively to be the class of languages:
$$
\mathcal{P}=\bigcup_{k\in\mathbb{N}_1}\text{DTIME}[n^k]
$$ 
and 
$$
\mathcal{NP}=\bigcup_{k\in\mathbb{N}_1}\text{NTIME}[n^k]. 
$$
      
The above definitions of complexity classes $\mathcal{P}$ and $\mathcal{NP}$ look a bit different from the official standard definitions in \cite{Coo00}. However, we will show in Appendix \ref{sec:appendix} that they are in fact equivalent.

With respect to the time complexity of a $k$-tape nondeterministic (respectively, deterministic) Turing machine and that of a single-tape nondeterministic (respectively, deterministic) Turing machine, we have the following useful lemma and corollary, extracted from \cite{AHU74} (see Lemma 10.1 and Corollary 1 to Lemma 10.1 in \cite{AHU74}), which play important roles in the following context:

\begin{lemma}[Lemma 10.1 in \cite{AHU74}] \footnote{ The deterministic version of this theorem appeared in \cite{HS65} for the first time (Theorem 6 in \cite{HS65}), and one can extend its proof to show the nondeterministic version of this lemma.}
\label{lemma1} 
If $L$ is accepted by a $k$-tape nondeterministic $T(n)$ time-bounded Turing machine, then $L$ is accepted by a single-tape nondeterministic $O(T^2(n))$ time-bounded Turing machine.\Q.E.D
\end{lemma}

\vskip 0.3 cm

The deterministic version of the above lemma is as follows:

\begin{corollary}[Corollary 1 in \cite{AHU74} to Lemma 10.1; see also Theorem 6 in \cite{HS65} and Theorem 2.1 in \cite{Pap94}] \footnote{ This corollary immediately follows from Lemma \ref{lemma1}, since a $k$-tape deterministic Turing machine is a special kind of $k$-tape nondeterministic Turing machine.}
\label{corollary4}
If $L$ is accepted by a $k$-tape deterministic $T(n)$ time-bounded Turing machine, then $L$ is accepted by a single-tape deterministic $O(T^2(n))$ time-bounded Turing machine.\Q.E.D
\end{corollary}

\vskip 0.3 cm

The following theorem about efficient simulation is needed a few times, and its proof is present in \cite{HS66} (see also \cite{AB09}).

\begin{lemma}[\cite{AB09}, Cf. \cite{HS66}]
\label{lemma2}
There exists a Turing machine $U$ such that for every $x,\alpha\in\{0,1\}^*$, $U(x,\alpha)=M_{\alpha}(x)$, where $M_{\alpha}$ denotes the Turing machine represented by $\alpha$. Moreover, if $M_{\alpha}$ halts on input $x$ within $T(|x|)$ steps, then $U(x,\alpha)$ halts within $cT(|x|)\log T(|x|)$ steps,\footnote{ In this work, $\log n$ stands for $\log_2 n$.} where $c$ is a constant independent of $|x|$ and depending only on $M_{\alpha}$'s alphabet size, number of tapes, and number of states.\Q.E.D
\end{lemma}

For a complexity class $\mathcal{C}$, its complement is denoted by ${\rm co}\mathcal{C}$ (see \cite{Pap94}), i.e., 
$$
{\rm co}\mathcal{C}=\{\overline{L}\,|\,L\in\mathcal{C}\}, 
$$
where $L$ is a decision problem, and $\overline{L}$ is the complement of $L$. For example, ${\rm co}\mathcal{P}$ is the complement of $\mathcal{P}$, and ${\rm co}\mathcal{NP}$ is the complement of $\mathcal{NP}$. Note that the complement of a decision problem $L$ is defined as the decision problem whose answer is ``{\em yes}" whenever the input is a ``{\em no}" input of $L$, and vice versa.

Finally, more information and premise lemmas will be given along the way to prove our main results.
    
\vskip 0.3 cm
\section{Enumeration of All Polynomial-Time Deterministic Turing Machines}
\label{sec:enumeration}
\vskip 0.3 cm

We should first clearly clarify the formal definition of polynomial-time deterministic Turing machines before coming to the point.

\begin{definition}[Cf. \cite{Coo00}]
\label{definition4}
Formally, a polynomial-time deterministic Turing machine is a deterministic Turing machine $M$ such that there exists $k\in\mathbb{N}_1$, for all input $x$ of length $|x|$, $M(x)$ will halt within $|x|^k+k$ steps.
\end{definition}
     
If a polynomial-time Turing machine runs at most $|x|^k$ steps for any input $x$, then we often say that it runs within time $O(n^{k-1})$ rather than $O(n^k)$ in the following context. 
     
For the purpose here, we should represent a polynomial-time deterministic Turing machine by a tuple of $(M,k)$ where $M$ is the polynomial-time deterministic Turing machine itself, and $k$ is the unique minimal degree of some polynomial $n^k+k$ such that for any input $x$ of length $n$, $M(x)$ will halt within $n^k+k$ steps. In the following context, we also call $k$ the order of the polynomial-time deterministic Turing machine represented by the tuple $(M,k)$.

\vskip 0.3 cm
\begin{remark}
\label{remark1}
 Obviously, in the above definition, given a polynomial-time deterministic Turing machine $(M,k)$, for any input $x$ of length $|x|$, $M(x)$ will halt within $O(|x|^{k+i})$ steps, where $i\geq 0$. But there exists some input $y$ of length $|y|$ such that $M(y)$ does not halt within $O(|y|^{k-1})$ steps. Further, the tuple representation $(M,k)$ of a deterministic $n^k+k$ time-bounded Turing machine $M$ has some advantages in this work. Namely, it clearly indicates that $M$ is a polynomial-time deterministic Turing machine and that the order of $(M,k)$ is $k$, from which we can easily recover the minimal polynomial $n^k+k$.
 \end{remark}
 \vskip 0.3 cm
      
By Corollary \ref{corollary4}, we can restrict ourselves to single-tape deterministic Turing machines. So, in the following context, by polynomial-time deterministic Turing machines, we mean single-tape polynomial-time deterministic Turing machines.

To obtain our main results, we need to {\em enumerate} the polynomial-time deterministic Turing machines so that for each positive integer $i$ there is a unique tuple of $(M,k)$ associated with $i$ (i.e., to define a function from $\mathbb{N}_1$ to the set of all polynomial-time deterministic Turing machines $\{(M,k)\}$ such that the function is surjective \footnote{ There are a variety of ways to enumerate all polynomial-time deterministic Turing machines. For instance, see proof of Theorem 14.1 in \cite{Pap94}, p. 330, or see \cite{Lad75}.}) such that we can refer to the $j$-th polynomial-time deterministic Turing machine.

To achieve our goals, we first use the method presented in \cite{AHU74}, p. 407, to encode a single-tape deterministic Turing machine into an integer.\footnote{ The way to enumerate deterministic Turing machines in \cite{AHU74} is basically the same as that of Turing \cite{Tur37} in principle, i.e., to define a mapping from Turing machines to positive integers, but with different specific details. See \cite{Tur37}, Section of {\it Enumeration of computable sequences}.}

Without loss of generality, we can make the following assumptions about the representation of a single-tape deterministic Turing machine with input alphabet $\{0,1\}$ because that will be all we need:
\begin{enumerate}
\item {The states are named $$q_1,q_2,\cdots,q_s$$ for some $s$, with $q_1$ the initial state and $q_s$ the accepting state.}
\item {The input alphabet is $\{0,1\}$.}
\item {The tape alphabet is $$\{X_1,X_2,\cdots,X_t\}$$ for some $t$, where $X_1=\mathbbm{b}$, $X_2=0$, and $X_3=1$.}
\item {The next-move function $\delta$ is a list of quintuples of the form $$(q_i,X_j,q_k,X_l,D_m),$$ meaning that $$\delta(q_i,X_j)=(q_k,X_l,D_m),$$ and $D_m$ is the direction, $L$, $R$, or $S$, if $m=1,2$, or $3$, respectively. We assume this quintuple is encoded by the string $$10^i10^j10^k10^l10^m1.$$}
\item {The deterministic Turing machine itself is encoded by concatenating in any order the codes for each of the quintuples in its next-move function. Additional $1$'s may be prefixed to the string if desired. The result will be some string of $0$'s and $1$'s, beginning with $1$, which we can interpret as an integer.}
\end{enumerate}

Next, we encode the order of $(M,k)$ to be $10^k1$ so that the tuple $(M,k)$ should be the concatenation of the binary string representing $M$ itself followed by the order $10^k1$. Now the tuple $(M,k)$ is encoded as a binary string, which can be explained as an integer. 
    
Any integer that cannot be decoded is deemed to represent the trivial polynomial-time deterministic Turing machine with an empty next-move function by this encoding. Every polynomial-time deterministic Turing machine will appear infinitely often in the enumeration since, given a polynomial-time deterministic Turing machine, we can prefix $1$'s at will to find larger and larger integers representing the same set of the polynomial-time deterministic Turing machine of $(M,k)$. We denote such a set of the polynomial-time deterministic Turing machine by $\widehat{M}_j$, where $j$ is the integer representing $(M,k)$. The reader will easily get that we have defined a surjective function $e$ from $\mathbb{N}_1$ to the set $\{(M,k)\}$ of all polynomial-time deterministic Turing machines, which is consistent with Definition \ref{definition1}. 
    
Furthermore, we in fact have defined a $(1,1)$ correspondence between the set $\{(M,k)\}$ of all polynomial-time deterministic Turing machines and $\mathbb{N}_1$ if any integer that cannot be decoded is deemed to represent the trivial polynomial-time deterministic Turing machine, from which we have reached the similar case to p. 241 of \cite{Tur37}, i.e., the set $\{(M,k)\}$ of all polynomial-time deterministic Turing machines is therefore enumerable.
    
\vskip 0.3 cm
\begin{remark}
One of the conveniences of tuple representation $(M,k)$ for a polynomial-time deterministic Turing machine in this way is, of course, to conveniently control the running time of the universal nondeterministic Turing machine $M_0$ constructed in Theorem \ref{theorem4} in Section \ref{sec:diagonalization1} below, so that it facilitates our analysis of the time complexity of $M_0$, i.e., to easily show the fact of Theorem \ref{theorem5}.
\end{remark}
\vskip 0.3 cm

Finally, we remark that the enumeration of all polynomial-time deterministic Turing machines also gives an enumeration of all languages in $\mathcal{P}$ (with languages appearing multiple times). In particular, we have the following theorem:

\begin{theorem}\footnote{ This theorem is somewhat redundant because an enumeration of the set $\{(M,k)\}$ is a surjective function $$e:\mathbb{N}_1\rightarrow\{(M,k)\},$$ so that for any element $(M,k)$, there is an $i\in\mathbb{N}_1$ such that $$e(i) = (M,k).$$ For simplicity, sometimes we just let $P$ denote the set $\{(M,k)\}$ of polynomial-time deterministic Turing machines.}
\label{theorem3}
All of the polynomial-time deterministic Turing machines are in the above enumeration $e$. In other words, the set $\{(M,k)\}$ of all polynomial-time deterministic Turing machines is enumerable. \Q.E.D
\end{theorem}

\vskip 0.3 cm
\begin{remark}
\label{remark2}
There is another way to {\em enumerate} all polynomial-time deterministic Turing machines without encoding the order of polynomial-time deterministic Turing machines into their representation. To do so, we need the {\em Cantor pairing function}: $$\pi:\mathbb{N}\times\mathbb{N}\rightarrow\mathbb{N}$$ 
defined by 
$$\pi(k_1,k_2):=\frac{1}{2}(k_1+k_2)(k_1+k_2+1)+k_2,$$
where $k_1,k_2\in\mathbb{N}$. Since the Cantor pairing function (see Figure \ref{2} below, which is from \cite{A2}) is invertible (see \cite{A2}), it is a bijection between $\mathbb{N}\times\mathbb{N}$ and $\mathbb{N}$. As we have shown that any polynomial-time deterministic Turing machine itself is an integer, we can place any polynomial-time deterministic Turing machine $M$ and its order $k$ in the tuple $(M,k)$ and use the Cantor pairing function to map the tuple $(M,k)$ to an integer in $\mathbb{N}_1$. Recall that there is a bijection between $\mathbb{N}$ and $\mathbb{N}_1$. Obviously, by Definition \ref{definition1}, the inverse of such a Cantor pairing function is an enumeration of the set $\{(M,k)\}$\footnote{ Thus, the set $\mathbb{N}_1\times\mathbb{N}_1$ is enumerable or countable.} of all polynomial-time deterministic Turing machines.
\end{remark}

\vskip 0.3 cm
\begin{figure}[htb]
     \center{\includegraphics[width=12.5cm]{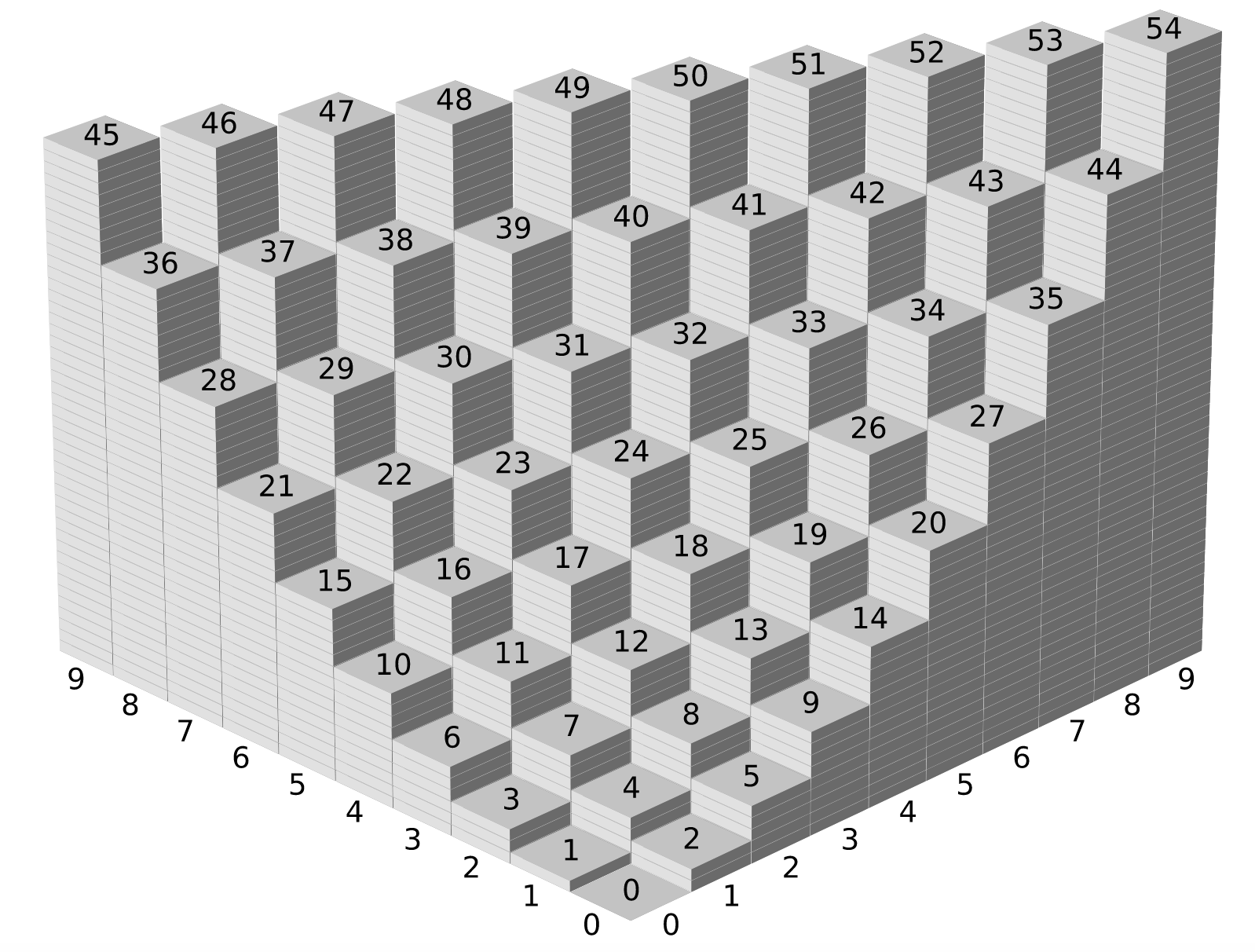}}
     \caption{\label{2}Cantor pairing function}
\end{figure}
\vskip 0.1 cm
   
\vskip 0.3 cm
\section{Diagonalization against All Polynomial-Time Deterministic Turing Machines}
\label{sec:diagonalization1}
\vskip 0.3 cm

We can now design a four-tape nondeterministic Turing machine $M_0$ that treats its input string $x$ both as an encoding of a tuple $(M,k)$ and also as the input to the polynomial-time deterministic Turing machine $M$. One of the capabilities possessed by $M_0$ is the ability to simulate a deterministic Turing machine, given its specification. By Lemma \ref{lemma2}, the simulation can be done within $$O(T(n)\log T(n))$$ steps, so we shall have $M_0$ determine whether the deterministic $n^k+k$ time-bounded Turing machine $(M,k)$ accepts the input $x$ without using more than $O(T(n)\log T(n))$ steps, where $$T(n)=n^k+k.$$ If $M$ accepts $x$ within time $n^k+k$, then $M_0$ does not. Otherwise, $M_0$ accepts $x$. Thus, for all $i$, $M_0$ disagrees with the behavior of the $i$-th deterministic $n^k+k$ time-bounded Turing machine in the enumeration $e$ on that input $x$. Concretely, we are going to show the following:

\begin{theorem}
\label{theorem4}
There exists a language $L_d$ accepted by a universal nondeterministic Turing machine $M_0$ but by no polynomial-time deterministic Turing machines.
\end{theorem}

\begin{proof}
Let $M_0$ be a four-tape nondeterministic Turing machine that operates as follows on an input string $x$ of length $n$.
\begin{enumerate}
\item{$M_0$ decodes the tuple encoded by $x$. If $x$ is not the encoding of some polynomial-time deterministic Turing machine $\widehat{M}_j$ for some $j$, then GOTO $5$; else determine $t$, the number of tape symbols used by $\widehat{M}_j$; $s$, its number of states; and $k$, its order. The third tape of $M_0$ can be used as ``scratch" memory to calculate $t$.}
\item{Then $M_0$ lays off on its second tape $n$ blocks of $$\lceil\log t\rceil$$ cells each, the blocks being separated by a single cell holding a marker $\#$, i.e., there are $$(1+\lceil\log t\rceil)n$$ cells in all. Each tape symbol occurring in a cell of $\widehat{M}_j$'s tape will be encoded as a binary number in the corresponding block of the second tape of $M_0$. Initially, $M_0$ places $\widehat{M}_j$'s input, in binary coded form, in the blocks of tape $2$, filling the unused blocks with the code for the blank.}
\item{On tape $3$, $M_0$ sets up a block of $$\lceil(k+1)\log n\rceil$$ cells, initialized to all $0$'s. Tape $3$ is used as a counter to count up to \[n^{k+1}.\footnote{ Assume that $\widehat{M}_j$ is a deterministic $T(n)$ time-bounded Turing machine where $$T(n)=n^k+k,$$ then by Lemma \ref{lemma2}, the simulation by $M_0$ can be done within time $T(n)\log T(n)$, which is less than $n^{k+1}$, so we set the counter to count up to $n^{k+1}$.}\]}
\item{By using nondeterminism,\footnote{ When $M_0$ simulates a deterministic Turing machine, the behavior of $M_0$ is somewhat deterministic because there are no nondeterministic choices in a deterministic Turing machine.} $M_0$ simulates $\widehat{M}_j$, using tape $1$, its input tape, to determine the moves of $\widehat{M}_j$ and using tape $2$ to simulate the tape of $\widehat{M}_j$. The moves of $\widehat{M}_j$ are counted in binary in the block of tape $3$, and tape $4$ is used to hold the state of $\widehat{M}_j$. If $\widehat{M}_j$ accepts, then $M_0$ halts without accepting. $M_0$ accepts if $\widehat{M}_j$ halts without accepting, or if the counter on tape $3$ overflows, $M_0$ halts without accepting.}\label{item4}
\item{Since $x$ is not an encoding of some deterministic Turing machine. Then $M_0$ sets up a block of $$\lceil 2\times\log n\rceil$$ cells on tape $3$, initialized to all $0$'s. Tape $3$ is used as a counter to count up to $$n^2.$$ By using its nondeterministic choices, $M_0$ moves as per the path described by $x$. The moves of $M_0$ are counted in binary in the block of tape $3$. If the counter on tape $3$ overflows, then $M_0$ halts. $M_0$ accepts $x$ if and only if there is a computation path from the start state of $M_0$ leading to the accept state and the total number of moves can not exceed $$n^2$$ steps, so it is within $O(n)$. Note that the number of $2$ in $$\lceil 2\times\log n\rceil$$ is fixed, i.e., it is a default setting (or it is by default).}
\end{enumerate}

The nondeterministic Turing machine $M_0$ described above is of time complexity, say $S$, which is currently unknown. According to Lemma \ref{lemma1}, $M_0$ is equivalent to a single-tape nondeterministic $$O(S^2)$$ time-bounded Turing machine, and it of course accepts some language $L_d$.

Suppose now $L_d$ were accepted by some, say, the $i$-th deterministic Turing machine in the enumeration $e$, which is a deterministic $$T(n)=n^k+k$$ time-bounded Turing machine $\widehat{M}_i$. Then by Corollary \ref{corollary4}, we may assume that $\widehat{M}_i$ is a single-tape polynomial-time deterministic Turing machine. Let $\widehat{M}_i$ have $s$ states and $t$ tape symbols. Since $\widehat{M_i}$ appears infinitely often\footnote{ We know that we may prefix $1$s at will to find larger and larger integers representing the same set of quintuples of the same deterministic Turing machine $M_i$; thus, there are infinitely many binary strings of sufficient length that represent deterministic Turing machine $M_i$.} in the enumeration $e$, $M_0$ was to set the counter of tape $3$ to count up to $n^{k+1}$, and

\begin{align*}
\lim_{n\rightarrow\infty}&\frac{T(n)\log T(n)}{n^{k+1}}\\
 =&\lim_{n\rightarrow\infty}\frac{(n^k+k)\log(n^k+k)}{n^{k+1}}\\
 =&\lim_{n\rightarrow\infty}\Big(\frac{n^k\log(n^k+k)}{n^{k+1}}+\frac{k\log(n^k+k)}{n^{k+1}}\Big)\\
 =&0\\
 <&1.
\end{align*}

So, there exists an $N_0>0$ such that for any $N\geq N_0$, $$T(N)\log T(N)<N^{k+1},$$ which implies that for a sufficiently long $w$, say $|w|\geq N_0$, and $M_w$ denoted by such $w$ is $\widehat{M}_i$, we have $$T(|w|)\log T(|w|)<|w|^{k+1}.$$

Thus, on input $w$, $M_0$ has sufficient time to simulate $M_w$ and accepts if and only if $M_w$ rejects.\footnote{ In the simulation of a polynomial-time deterministic Turing machine, $M_0$ only turns itself off mandatorily when the counter on tape $3$ overflows, i.e., its counter $$\geq |w|^{k+1}, $$ where $w$ is the input.} But we assumed that $\widehat{M}_i$ accepted $L_d$, i.e., $\widehat{M}_i$ agreed with $M_0$ on all inputs. We thus conclude that $\widehat{M}_i$ does not exist, from which it immediately follows $$L_d\not\in \mathcal{P}. $$ This completes the proof.
\end{proof}
   
\vskip 0.1 cm  
We need to remark more about item (\ref{item4}) in the proof of Theorem \ref{theorem4}. Because the universal nondeterministic Turing machine $M_0$ diagonalizes against all polynomial-time deterministic Turing machines, it can flip the answer immediately when the simulation ends due to $$\mathcal{P}={\rm co}\mathcal{P}. $$ If it is the case that $M_0$ diagonalizes against a list of polynomial-time nondeterministic Turing machines, then other techniques must be presented, since we do not know whether the following relationship $$\mathcal{NP}={\rm co}\mathcal{NP}$$ holds or not. Currently, we completely have no idea about how to prove or disprove this important relationship.

\vskip 0.3 cm
\begin{remark}
\label{remark3}
More simply, the reader can regard the universal nondeterministic Turing machine $M_0$ as a combination of a deterministic universal Turing machine $\mathcal{U}$ and a nondeterministic Turing machine $\mathcal{H}$, the similar construction of which appeared in Turing \cite{Tur37} for the first time (see p. 247, \cite{Tur37}).

In fact, we can design our universal nondeterministic Turing machine $M_0$ to be more complicated. For example, since we can encode any polynomial-time nondeterministic Turing machine into a binary string representing itself followed by its order $10^k1$ (see \cite{Lin21b}), the input to $M_0$ can be classified into three types: If the input is a polynomial-time deterministic Turing machine, then $M_0$ does the work specified in the proof of Theorem \ref{theorem4} except item (5); if the input is a polynomial-time nondeterministic Turing machine $N$ (of time complexity $n^k+k$), then $M_0$ can set the counter of tape $3$ to count up to $$n^{k+1},$$ then simulate $N$ nondeterministically in time $n^{k+1}$ and output its answer, i.e., accepting if $N$ accepts and rejecting if $N$ rejects; otherwise, $M_0$ rejects the input. Note that such a design does not change $M_0$'s time complexity.
     
In general, the diagonalization techniques used in the proof of Theorem \ref{theorem4} can not directly apply to diagonalizing against a collection of nondeterministic $$T(n)$$ time-bounded Turing machines within time $$L(n),$$ where $$\lim_{n\rightarrow\infty}\frac{T(n)}{L(n)} = 0.$$ Because a nondeterministic Turing machine that runs in time $$O(T(n))$$ may have $$2^{O(T(n))}$$ branches, it is unclear how to determine whether it accepts and then flips the answer in time $$O(L(n)),$$ assuming that $$L(n)<2^{O(T(n))}. $$ That is, we do not know whether $$\mathcal{NP}={\rm co}\mathcal{NP},$$ as observed by Cook \cite{Coo73}. Thus, in his work \cite{Coo73}, Cook uses different techniques and then, by ``indirect" diagonalization, which is somewhat complicated (e.g., the ingredient of technique from \cite{Iba72} with other complicated techniques), to show a nondeterministic time hierarchy theorem \cite{Coo73} (it has to say that there exist other techniques such as {\em lazy diagonalization} that are also capable of showing nondeterministic time hierarchy theorems; see e.g., \cite{AB09,For00,FS07,Zak83}. More surprisingly, Fortnow \cite{FS17} developed a much more elegant and simple style of diagonalization to show nondeterministic time hierarchy). But fortunately, we diagonalize against all polynomial-time deterministic Turing machines rather than polynomial-time nondeterministic Turing machines in Theorem \ref{theorem4} and its proof, so our nondeterministic Turing machine $M_0$ can flip the answer immediately when the simulation ends.
\end{remark}
\vskip 0.3 cm

Next, we are going to show that the universal nondeterministic Turing machine $M_0$ runs within time $O(n^k)$ for any $k\in\mathbb{N}_1$:

\begin{theorem}
\label{theorem5}
The universal nondeterministic Turing machine $M_0$ constructed in proof of Theorem \ref{theorem4} runs within time $O(n^k)$ for any $k\in\mathbb{N}_1$.
\end{theorem}

\begin{proof}
The simplest way to show the theorem is to prove that for any input $w$ to $M_0$, there is a corresponding positive integer $i_w\in\mathbb{N}_1$ such that $M_0$ runs at most $$|w|^{i_w+1}$$ steps, which can be done as follows. 
    
On the one hand, if the input $x$ encodes a deterministic $n^k+k$ time-bounded Turing machine, then $M_0$ turns itself off mandatorily within $$|x|^{k+1}$$ steps by the construction, so the corresponding integer $i_x$ is $k$ in this case (i.e., $i_x=k$). This holds true for all polynomial-time deterministic Turing machines as input with $k$ to be the order of that corresponding polynomial-time deterministic Turing machine. 
    
But on the other hand, if the input $x$ does not encode some polynomial-time deterministic Turing machine, then the running time of $M_0$ is within time $$O(|x|)$$ because $M_0$ must turn itself off when the counter exceeds $$|x|^2$$ steps by the construction, so the corresponding integer $i_x$ is $1$ in this case (i.e., $i_x=1$). In both cases we have shown that for any input $w$ to $M_0$, there is a corresponding positive integer $i_w\in\mathbb{N}_1$ such that $M_0$ runs at most $|w|^{i_w+1}$ steps. So $M_0$ is a nondeterministic $$ S(n)=\max\{n^k,n\}$$ time-bounded Turing machine for any $k\in\mathbb{N}_1$. Thus, $M_0$ is a nondeterministic $$O(n^k)$$ time-bounded Turing machine for any $k\in\mathbb{N}_1$. By Lemma \ref{lemma1}, there is a single-tape nondeterministic Turing machine $M'$ equivalent to $M_0$, and $M'$ runs within time $$O(S(n)^2)=O(n^{2k})$$ for any $k\in\mathbb{N}_1$.
\end{proof}

\vskip 0.3 cm
\section{Proof of $L_d\in\mathcal{NP}$}
\label{sec:proof_ldinnp}
\vskip 0.3 cm

At the moment, since the exponent $k$ in $O(n^k)$, which is the running time of $M_0$, can be any integer in $\mathbb{N}_1$, it is not so transparent for some readers that the language $L_d$ accepted by $M_0$ running in $O(n^k)$ for all $k\in\mathbb{N}_1$ is in $\mathcal{NP}$. For example, see \cite{For21}, in which complexity theorist Fortnow claims that for $M_0$ to run in polynomial time, it must run in time $O(n^c)$ for a fixed $c$. But to diagonalize all polynomial-time deterministic Turing machines, then $M_0$ needs time $O(n^k)$ for all $k\in\mathbb{N}_1$, including $k>c$. He hence asserts that the author made a common mistake, and this mistake is not fixable due to the result of \cite{BGS75}. In view of the above, at this moment in time the most notable difficulty is that there is no fixed constant $c\in\mathbb{N}_1$ such that $M_0$ runs within time $O(n^c)$. We had to overcome such obstacles to show the fact of the theorem below, and admittedly, overcoming such obstacles requires sophisticated observation from novel perspectives (however, there is a simple and fast way to get the intuition that $L_d\in\mathcal{NP}$, see {\em footnote} \ref{footnote26}, although the {\em footnote} \ref{footnote26} is not considered a rigorous mathematical proof):

\begin{theorem}\footnote{ To be more accurate, we should discuss this theorem with a single-tape nondeterministic time-bounded Turing machine $\widehat{M}_0$ that is of time complexity $$O(n^{2k})$$ for any $k\in\mathbb{N}_1$ accepting the language $L_d$. But the final outcome is the same as $M_0$.}
\label{theorem6}
The language $L_d$ is in $\mathcal{NP}$, where $L_d$ is accepted by $M_0$, which runs within time $O(n^k)$ for any $k\in\mathbb{N}_1$.
\end{theorem}

\begin{proof}
Let us first define the family of languages $$\{L_d^i\}_{i\in\mathbb{N}_1}$$ as follows:
\begin{align*}
L_d^i\overset{{\rm def}}{=}&\text{ language accepted by $M_0$ running within time $O(n^i)$ for a fixed $i\in\mathbb{N}_1$.}\\
&\text{ That is, $M_0$ turns itself off mandatorily when its moves made by itself }\\
&\text{ during the computation exceed $n^{i+1}$ steps.}
\end{align*}
 
Note that the above definition of language $L_d^i$ technically can be done by adding a new tape to $M_0$ as a counter to count up to $$n^{i+1}$$ for a fixed $i\in\mathbb{N}_1$, meaning that $M_0$ turns itself off when the counter of tape $3$ exceeds $$n^{k+1}$$ or the counter of the newly added tape exceeds $$n^{i+1}.$$  Obviously, for each $i\in\mathbb{N}_1$, $L_d^i$ is a truncation of $L_d$. 

Then by the construction of $M_0$, namely, for an arbitrary input $w$ to $M_0$, there is a corresponding integer $$i_w\in\mathbb{N}_1$$ such that $M_0$ runs at most $$|w|^{i_w+1}$$ steps (in other words, $M_0$ runs at most $n^{i+1}$ steps for any $i\in\mathbb{N}_1$ where $n$ is the length of the input; see Theorem \ref{theorem5} above), we have
\begin{equation}
\label{eq1}
\begin{split}
L_d=&\bigcup_{i\in\mathbb{N}_1}L_d^i.
\end{split}
\end{equation}

Furthermore, $$L_d^i\subseteq L_d^{i+1},\quad\text{for each fixed $i\in\mathbb{N}_1$,}$$ since for any word $w\in L_d^i$ accepted by $M_0$ within $O(n^i)$ steps, it surely can be accepted by $M_0$ within $O(n^{i+1})$ steps, i.e.,
\[w\in L_d^{i+1}.\footnote{ By Cook's proof \cite{Coo71}, it is clear that $L_d^i$ can be reduced to {\em Satisfiability} (SAT) for each fixed $i\in\mathbb{N}_1$.} \]

This gives that for any fixed $i\in\mathbb{N}_1$,
\begin{equation}
\label{eq2}
\begin{split}
  L_d^1\subseteq L_d^2\subseteq\cdots\subseteq L_d^i\subseteq L_d^{i+1}\subseteq\cdots
\end{split}
\end{equation}

Note further that for any fixed $i\in\mathbb{N}_1$, $L_d^i$ is the language accepted by the nondeterministic Turing machine $M_0$ within time $O(n^i)$, i.e., at most $n^{i+1}$ steps,\footnote{ When $M_0$ turns itself off mandatorily when its moves made by itself during the computation exceed $$n^{i+1}$$ steps, it is a polynomial-time nondeterministic Turing machine, i.e., it is a nondeterministic $$n^{i+1}+(i+1)$$ time-bounded Turing machine.} we thus obtain

\begin{equation}
       \label{eq3}
       \begin{split}
         L_d^i\in\text{NTIME}[n^i]\subseteq\mathcal{NP},\quad\text{for any fixed $i\in\mathbb{N}_1$}.
       \end{split}
\end{equation}

Now, (\ref{eq1}) together with (\ref{eq2}) and (\ref{eq3}) easily implies $$L_d\in \mathcal{NP},$$ as required.
\end{proof}

\vskip 0.3 cm
\subsection{The $2$nd Simpler Proof}
\vskip 0.3 cm
    
In fact, we can prove Theorem \ref{theorem6} by contradiction, which is much simpler. To do so, after obtaining the relations (\ref{eq1}) and (\ref{eq2}), we can assume that $$L_d\not\in\mathcal{NP}, $$ then there must exist at least a fixed $i\in\mathbb{N}_1$ such that $$L_d^i\not\in\mathcal{NP}. $$ But by definition, $L_d^i$ is the language accepted by the nondeterministic Turing machine $M_0$ running at most $$n^{i+1}$$ steps (or, within time $O(n^i)$), which clearly is a contradiction. We thus can claim that such an $i$ can not be found. Equivalently, $$L_d^i\in\mathcal{NP}$$ for all $i\in\mathbb{N}_1$, which, together with the relations (\ref{eq1}) and (\ref{eq2}), further implies $$L_d\in\mathcal{NP}, $$ as required.\Q.E.D

\vskip 0.3 cm
\begin{remark}
\label{remark4}
Comparing the above two proofs of $L_d\in\mathcal{NP}$, we are in favor of the second one because it is more concise and simpler than the first proof. 
       
In addition, when an earlier version of this manuscript was previously submitted to a journal, we got the feedback that our simulating machine $M_0$ does not run in polynomial time, i.e., it asserts that $$L_d\not\in\mathcal{NP}$$. With regard to those comments, we present Theorem \ref{theorem6} most clearly and most rigorously to show that $$L_d\in\mathcal{NP}. $$

So one can carefully examine the proof of Theorem \ref{theorem6} and then point out whether the proof is correct or not, and further, she/he can point out why our proof of Theorem \ref{theorem6} is wrong.\footnote{ One can try to present counter-examples to the proof of Theorem \ref{theorem6} or to prove that all nondeterministic Turing machines accepting the language $L_d$ require $$\Omega(c^n)$$ steps, for some constant $c>1$.}  
       
Now, let us return to such a question. Although we have shown in Theorem \ref{theorem6} that $$L_d\in\mathcal{NP}$$ mathematically, some readers are left with such a question: Can we find a fixed constant $$t\in\mathbb{N}_1$$ such that the nondeterministic Turing machine $M_0$ runs within polynomial time $$n^t+t\, ? $$ The answer depends on whether we can answer the following question: Let $${\rm DPTMs}=\{T_1,T_2,\cdots\}$$ be the set of all polynomial-time deterministic Turing machines, and let $${\rm order}(T_i)$$ be the order of $T_i$, i.e., ${\rm order}(T_i)$ is the degree of the minimal polynomial of $T_i$. For example, if $T_i$ is a deterministic $n^l+l$ time-bounded Turing machine, then $${\rm order}(T_i)=l. $$ Let $$m=\max\{{\rm order}(T_1), {\rm order}(T_2),\cdots\}, $$ then we can say $$n^{m+1}+(m+1)$$ is the minimal polynomial of $M_0$. But here comes the key question, i.e., can we find such a fixed constant $z$ in $\mathbb{N}_1$ such that $$z=m\,? $$
\end{remark}

\vskip 0.3 cm
\subsection{Proof of Theorem \ref{theorem1}}
\label{sec:proofofmaintheorem1}
\vskip 0.3 cm

Now we are at the point to give the proof of Theorem \ref{theorem1}:

\vskip 0.2 cm

\indent{\it Proof}. It is obvious that Theorem \ref{theorem1} is an immediate consequence of Theorem \ref{theorem4} and Theorem \ref{theorem6}. \Q.E.D

\vskip 0.3 cm
\begin{remark}
\label{remark5}

Originally, we call $M_0$ a polynomial-time nondeterministic Turing machine. But complexity theorist Lance Fortnow \cite{For21}\footnote{ We received valuable criticisms \cite{For21} from Lance Fortnow shortly after posting the manuscript of this work on the arXiv.} argues that $M_0$ does not run in polynomial time because it runs in time $O(n^k)$ for any $k\in\mathbb{N}_1$. He thinks the notion of a polynomial-time machine is a fixed mathematical definition, i.e., call a machine running in polynomial time iff the machine runs in time $O(n^c)$ for some fixed constant $c>0$. For this, see special thanks expressed in the acknowledgements section; see Section \ref{sec:acks}.

However, on the other hand, as we have shown in Theorem \ref{theorem6}, the language $L_d$ accepted by $M_0$ is indeed in $\mathcal{NP}$, which is sufficient for our discussions. In a nutshell, what we are interested in is constructing a language $$ L_d\not\in\mathcal{P}$$ but $$L_d\in\mathcal{NP}.$$ 

Hence, the distinction between $M_0$ running within time $O(n^c)$ for some fixed $c$ and $M_0$ running within time $O(n^k)$ for all $k\in\mathbb{N}_1$ is unimportant in our setting. In fact, there is no machine that can run within time $O(n^c)$ for some fixed $c>0$ accepting the $L_d$ because the mathematicians acknowledge that $\mathbb{N}_1$ is not bounded from above.\footnote{ In fact, we can imagine if $\mathbb{N}_1$ is bounded from above, i.e., $$i\leq b \text{ for each } i\in\mathbb{N}_1,$$ then $M_0$ is a nondeterministic $n^b+b$ time-bounded Turing machine (i.e., it runs within time $O(n^b)$) where $b$ is the ``upper-bound" of $\mathbb{N}_1$, from which we have the intuition that $L_d\in\mathcal{NP}$.\label{footnote26}}
\end{remark}
    
\vskip 0.3 cm
\section{A Language $L_s\in\mathcal{P}$ That's Similar to But Different from $L_d$}
\label{sec:dontsuspect}
\vskip 0.3 cm

If Theorem \ref{theorem6} were not proved or not discovered, some experts would not regard that $L_d\in\mathcal{NP}$. For example, see the opening descriptions of Section \ref{sec:proof_ldinnp}. But, we should emphasize that, even though $M_0$ does not run within time $O(n^c)$ for fixed $c>0$, $L_d$ is indeed in $\mathcal{NP}$. Let us take a further example that resembles that $L_d$ is in $\mathcal{NP}$, i.e., we are going to construct a language $L_s\in\mathcal{P}$ with the properties that it also needs a specific deterministic Turing machine running within time $O(n^m)$ for all $m\in\mathbb{N}_1$ to accept. Our next result looks like the following.
  
\vskip 0.3 cm
\subsection{Simulation of Polynomial-Time DTMs}
\label{simulation_polynomial_time_dtms}
\vskip 0.3 cm

\begin{theorem}
\label{theorem7}
There exists a language $L_s$ accepted by a universal deterministic Turing machine $M'_0$ being of time complexity $O(n^k)$ for any $k\in\mathbb{N}_1$.
\end{theorem}

\begin{proof}
Let $M'_0$ be a four-tape deterministic Turing machine that operates as follows on an input string $x$ of length $n$.
     \begin{enumerate}
       \item{$M'_0$ decodes the tuple encoded by $x$. If $x$ is not the encoding of some single-tape polynomial-time deterministic Turing machine $\widehat{M}_j$ for some $j$ then rejects; else determine $t$, the number of tape symbols used by $\widehat{M}_j$; $s$, its number of states; and $k$, its order. The third tape of $M'_0$ can be used as ``scratch" memory to calculate $t$.}
       \item{Then $M'_0$ lays off on its second tape $n$ blocks of $$\lceil\log t\rceil$$ cells each, the blocks being separated by a single cell holding a marker $\#$, i.e., there are $$(1+\lceil\log t\rceil)n$$ cells in all. Each tape symbol occurring in a cell of $\widehat{M}_j$'s tape will be encoded as a binary number in the corresponding block of the second tape of $M'_0$. Initially, $M'_0$ places $\widehat{M}_j$'s input, in binary coded form, in the blocks of tape $2$, filling the unused blocks with the code for the blank.}
       \item{On tape $3$, $M'_0$ sets up a block of $$\lceil(k+1)\log n\rceil$$ cells, initialized to all $0$'s. Tape $3$ is used as a counter to count up to $$n^{k+1}. $$}
       \item{$M'_0$ simulates $\widehat{M}_j$, using tape $1$, its input tape, to determine the moves of $\widehat{M}_j$ and using tape $2$ to simulate the tape of $\widehat{M}_j$. The moves of $\widehat{M}_j$ are counted in binary in the block of tape $3$, and tape $4$ is used to hold the state of $\widehat{M}_j$. If the counter on tape $3$ overflows, $M'_0$ turns itself off mandatorily and rejects; else $M'_0$ accepts if and only if $\widehat{M}_j$ accepts.}
     \end{enumerate}

By arguments similar to those in the proof of Theorem \ref{theorem5}, it is not hard to show that for any input $w$ to $M'_0$, there is a corresponding positive integer $i_w\in\mathbb{N}_1$ such that $M'_0$ runs at most $|w|^{i_w+1}$ steps, thus demonstrating that the deterministic Turing machine $M'_0$ described above is of time complexity $O(n^k)$ for any $k\in\mathbb{N}_1$ and, of course, accepts some language $L_s$. This completes the proof.
\end{proof}

\vskip 0.3 cm
\subsection{Proof of $L_s$ in $\mathcal{P}$}
\label{proof_ls_in_P}
\vskip 0.3 cm

Now the language $L_s$ can be proved in $\mathcal{P}$ by similar arguments in Theorem \ref{theorem6}, given as follows:
\begin{lemma}
\label{lemma3}
      $L_s\in\mathcal{P}$ where $L_s$ is accepted by $M_0'$ within time $O(n^k)$ for any $k\in\mathbb{N}_1$.
\end{lemma}

\begin{proof}
By Corollary \ref{corollary4}, the deterministic Turing machine $M'_0$ is equivalent to a single-tape deterministic Turing machine being of time complexity $O(n^{2k})$ for any $k\in\mathbb{N}_1$.

We first define the family of languages $$\{L_s^i\}_{i\in\mathbb{N}_1}$$ as follows:
\begin{align*}
      L_s^i\overset{{\rm def}}{=}&\text{ language accepted by $M_0'$ within $O(n^i)$ steps, i.e., $M_0'$ turns itself off }\\
      \quad& \text{ mandatorily when its moves made by itself during the computation }\\
      \quad& \text{exceed $n^{i+1}$ steps.}
\end{align*}

Similarly, the above definition of language $L_s^i$ technically can be done by adding a new tape to $M_0'$ as a counter to count up to $$n^{i+1}$$ for a fixed $i\in\mathbb{N}_1$, meaning that $M_0'$ turns itself off when the counter of tape $3$ exceeds $$n^{k+1}$$ or the counter of the newly added tape exceeds $$n^{i+1}.$$ 

Then by construction, i.e., since for any input $w$ to $M'_0$, there is a corresponding integer $i_w\in\mathbb{N}_1$ such that $M'_0$ runs at most $|w|^{i_w+1}$ steps, we thus have
\begin{equation}
       \label{eq4}
       \begin{split}
        L_s=&\bigcup_{i\in\mathbb{N}_1}L_s^i.
       \end{split}
\end{equation}

Furthermore, $$L_s^i\subseteq L_s^{i+1},\quad\text{for each fixed $i\in\mathbb{N}_1$.} $$
Since for any word $w\in L_s^i$ accepted by $M'_0$ within $O(n^i)$ steps, it surely can be accepted by $M'_0$ within $O(n^{i+1})$ steps, i.e., $$w\in L_s^{i+1}. $$

This gives that for any $i\in\mathbb{N}_1$,
\begin{equation}
       \label{eq5}
       \begin{split}
          L_s^1\subseteq L_s^2\subseteq\cdots\subseteq L_s^i\subseteq L_s^{i+1}\subseteq\cdots
       \end{split}
\end{equation}

Note further that for any fixed $i\in\mathbb{N}_1$, $L_s^i$ is the language accepted by $M'_0$ running within time $O(n^i)$. In other words, $L_s^i$ is accepted by a single-tape deterministic Turing machine $N_0$ of time complexity $O(n^{2i})$, which is equivalent to $M'_0$ running within time $O(n^i)$. We thus obtain that
\begin{equation}
      \label{eq6}
      \begin{split}
       L_s^i\in\text{DTIME}[n^{2i}]\subseteq\mathcal{P},\quad\text{for any fixed $i\in\mathbb{N}_1$}.
       \end{split}
\end{equation}

From (\ref{eq4}), (\ref{eq5}), and (\ref{eq6}), we deduce that $$L_s\in \mathcal{P},$$ as required.
\end{proof}

\vskip 0.3 cm
\subsection{Proof of Theorem \ref{theoremByProduct}}
\label{sec:proof_of_theorembyproduct}
\vskip 0.3 cm
   
Although the language $L_s$ is in $\mathcal{P}$ as shown above, we conjecture that there is no deterministic Turing machine running within time $O(n^c)$ for a fixed $c>0$ that can accept it. Of course, by our constructions of machines $M_0$ in Theorem \ref{theorem4} and $M_0'$ in Theorem \ref{theorem7}, it is easy to see that $L_d\neq L_s$. \footnote{ $L_d$ and $L_s$ denote diagonalization language and simulation language, respectively.}
   
\vskip 0.3 cm
Now the proof of Theorem \ref{theoremByProduct} can be made naturally as follows:
\vskip 0.1 cm
\indent {\em Proof of Theorem \ref{theoremByProduct}.} It clearly follows from Theorem \ref{theorem7} and Lemma \ref{lemma3}. The proof is completed. \Q.E.D
   
\vskip 0.3 cm
\begin{remark}
Can we now deny that the language $L_s$ is not in $\mathcal{P}$ due to the machine $M'_0$ accepting it running within time $$O(n^k)$$ for all $k\in\mathbb{N}_1$? Of course, we can not, because the language $L_s$ accepted by $M'_0$ is in $\mathcal{P}$ but not others.\footnote{ In fact, we can construct a universal deterministic Turing machine running within time $O(n^i)$ for any $i\in\mathbb{N}_1$ accepting the language $$L'_s=\{(\langle M_i\rangle,w)\,|\, w\in L(M_i)\},$$ where $\langle M_i\rangle$ is the binary string representation of some polynomial-time deterministic Turing machine taken from the set $\{(M,k)\}$, and $L(M_i)$ denotes the language accepted by $M_i$, and the tuple of $(\langle M_i\rangle,w)$ can be seen as a pairing function, which is similar to \cite{DK14}; see Proposition 1.17, p. 26 in \cite{DK14}.\label{footnote28}} Moreover, still denote the set of all the polynomial-time deterministic Turing machines by $\{(M,k)\}$, then $$n^{(m+1)}+(m+1)$$ is the polynomial of $M'_0$ where $$m=\max\{{\rm order}(T)\,|\,T\in\{(M,k)\}\},$$ if $m$ exists.
\end{remark}

\vskip 0.3 cm
\section{Breaking the ``Relativization Barrier"}
\label{sec:relativization_barrier}
\vskip 0.3 cm

After the preliminary manuscript was posted on the online preprint repository, the author received doubts from another enthusiastic reader about the problem discussed in this work, i.e., any method that leads to $\mathcal{P}\neq\mathcal{NP}$ must overcome the aforementioned ``Relativization Barrier." Although we have forgotten who emailed the author and the original email has flown away, we thank her/him for her/his interest in the author's manuscript. Here, we systematically discuss this question in response to that enthusiast.

The computation model we use in this section is the {\em query machines}, or the {\em oracle Turing machines}, which is an extension of the multi-tape Turing machine, i.e., Turing machines that are given access to a black box or ``oracle" that can magically solve the decision problem for some language $$O\subseteq\{0,1\}^*.$$ The machine has a special {\em oracle tape} on which it can write a string $$w\in\{0,1\}^*$$ and in one step gets an answer to a query of the form $$\text{``Is $w$ in $O$?"},$$ which can be repeated arbitrarily often with different queries. If $O$ is a difficult language (say, one that cannot be decided in polynomial time or is even undecidable), then this oracle gives the Turing machine additional power. We first quote its formal definition as follows:

\begin{definition}[Cf. \cite{AB09}, Deterministic Oracle Turing machines]
\label{definition5}
A {\em deterministic oracle Turing machine} is a deterministic Turing machine $M$ that has a special read-write tape we call $M$'s {\em oracle tape} and three special states $q_{query}$, $q_{yes}$, and $q_{no}$. To execute $M$, we specify in addition to the input a language $O\subseteq\{0,1\}^*$ that is used as the {\em oracle} for $M$. Whenever during the execution $M$ enters the state $q_{query}$, the machine moves into the state $q_{yes}$ if $w\in O$ and $q_{no}$ if $w\not\in O$, where $w$ denotes the contents of the special oracle tape. Note that, regardless of the choice of $O$, a membership query to $O$ counts only as a single computation step. If $M$ is an oracle machine, $O\subseteq\{0,1\}^*$ a language, and $x\in\{0,1\}^*$, then we denote the output of $M$ on input $x$ and with oracle $O$ by $M^O(x)$.
\end{definition}

The above Definition \ref{definition5} is for the {\em Deterministic Oracle Turing Machines}, and the {\em Nondeterministic Oracle Turing Machines} can be defined similarly.

If for every input $x$ of length $|x|$, all computations of $M^X$ end in less than or equal to $T(|x|)$ steps, then $M^X$ is said to be a $T(n)$ {\em time-bounded (nondeterministic) deterministic oracle Turing machine with oracle $X$}, or said to be {\em of time complexity $T(n)$}. The family of languages of deterministic time complexity $T(n)$ with oracle $X$ is denoted by $${\rm DTIME}^X[T(n)];$$ the family of languages of nondeterministic time complexity $T(n)$ with oracle $X$ is denoted by $${\rm NTIME}^X[T(n)]. $$ The notation $\mathcal{P}^X$ and $\mathcal{NP}^X$ is defined respectively to be the class of languages: $$\mathcal{P}^X=\bigcup_{k\in \mathbb{N}_1}{\rm DTIME}^X[n^k] $$ and $$\mathcal{NP}^X=\bigcup_{k\in\mathbb{N}_1}{\rm NTIME}^X[n^k]. $$

\vskip 0.3 cm
\subsection{Proof of Theorem \ref{theorem2}}
\label{sec:proofofmaintheorem}
\vskip 0.3 cm

In this subsection, we prove our main result of Theorem \ref{theorem2}. Before starting, we should first remind the reader that Theorem \ref{theorem3} is an important prerequisite for proving Theorem \ref{theorem1}:

We assume that {\bf (I)} the polynomial-time deterministic (nondeterministic) oracle Turing machines can be effectively represented as strings; and further suppose that {\bf (II)} there are universal nondeterministic oracle Turing machines that can simulate any other and flip the answer of other deterministic oracle Turing machines; and lastly suppose that {\bf (III)} the simulation can be done within time $$O(T(n)\log T(n)),$$ where $T(n)$ is the time complexity of the simulated deterministic oracle Turing machine (see {\it footnote} \ref{footnote6}), all of which are base assumptions and satisfy the aforementioned properties {\bf I} and {\bf II} given in Subsection \ref{subsec:main_results}.

Since the input tape and the working tape of an oracle Turing machine can be the same tape, we thus assume that the machines in the set $P^O$ of the polynomial-time deterministic oracle Turing machines with oracle $O$ are two-tape oracle Turing machines; one is the input tape, and the other is the oracle tape. 

\vskip 0.15 cm
Now, we are ready to prove Theorem \ref{theorem2}:

\vskip 0.15 cm
\indent{\em Proof of Theorem \ref{theorem2}.} We show Theorem \ref{theorem2} by contradiction. Suppose to the contrary that the set $P^O$ of all polynomial-time deterministic oracle Turing machines with oracle $O$ is enumerable, or in other words, the cardinality of $P^O$ is less than or equal to that of $\mathbb{N}_1$. Then we have an enumeration $$e:\mathbb{N}_1\rightarrow P^O.$$

Next, we construct a five-tape universal nondeterministic oracle Turing machine $M_0^O$ that operates as follows on an input string $x$ of length $n$:
\begin{enumerate}
       \item{ 
           $M^O_0$ decodes the tuple encoded by $x$. If $x$ is not the encoding of some polynomial-time deterministic oracle Turing machine $D^O_j$ for some $j$, then GOTO $6$; else determine $t$, the number of tape symbols used by $D^O_j$; $s$, its number of states; and $k$, its order.\footnote{ We suppose that the order of $D^O$, i.e., the minimal degree of some polynomial of $D^O$, is also encoded into $D^O$, similarly to the way presented in Section \ref{sec:enumeration}.} The third tape of $D^O_0$ can be used as ``scratch" memory to calculate $t$.}
       \item{ Then $D^O_0$ lays off on its second tape $n$ blocks of $$\lceil\log t\rceil$$ cells each, the blocks being separated by a single cell holding a marker $\#$, i.e., there are $$(1+\lceil\log t\rceil)n$$ cells in all. Each tape symbol occurring in a cell of $D^O_j$'s tape will be encoded as a binary number in the corresponding block of the second tape of $M^O_0$. Initially, $M^O_0$ places $D^O_j$'s input, in binary coded form, in the blocks of tape $2$, filling the unused blocks with the code for the blank.}
       \item{ On tape $3$, $M^O_0$ sets up a block of $$\lceil(k+1)\log n\rceil$$ cells, initialized to all $0$'s. Tape $3$ is used as a counter to count up to $$n^{k+1}. $$}
       \item{ On tape $4$, $M^O_0$ reads and writes the contents of the oracle tape of $D^O_j$. That is, tape $4$ is the oracle tape of $M^O_0$, which is used to simulate the oracle tape of $D^O_j$.}
       \item{ By using nondeterminism, $M^O_0$ simulates $D^O_j$, using tape $1$, its input tape, to determine the moves of $D^O_j$ and using tape $2$ to simulate the tape of $D^O_j$, further using tape $4$ to simulate the oracle tape of $D^O_j$. The moves of $D^O_j$ are counted in binary in the block of tape $3$, and tape $5$ is used to hold the state of $D^O_j$. If $D^O_j$ accepts, then $M^O_0$ halts without accepting. $M^O_0$ accepts if $D^O_j$ halts without accepting, or if the counter on tape $3$ overflows, $M^O_0$ halts without accepting.}
       \item{ Since $x$ is not an encoding of some polynomial-time deterministic oracle Turing machine with oracle $O$. Then $M^O_0$ sets up a block of $$\lceil 2\times\log n\rceil$$ cells on tape $3$, initialized to all $0$'s. Tape $3$ is used as a counter to count up to $$n^2.$$ By using its nondeterministic choices, $M^O_0$ moves as per the path given by $x$. The moves of $M^O_0$ are counted in binary in the block of tape $3$. If the counter on tape $3$ overflows, then $M^O_0$ halts. $M^O_0$ accepts $x$ if and only if there is a computation path from the start state of $M^O_0$ leading to the accept state and the total number of moves can not exceed $$n^2$$ steps, so it is within $$O(n)$$ steps. Note that the number of $2$ in $$\lceil 2\times\log n\rceil$$ is fixed, i.e., it is a default setting.}
\end{enumerate}

The nondeterministic oracle Turing machine $M^O_0$ described above is a nondeterministic oracle Turing machine that is of time complexity $$O(n^m)$$ for any $m\in\mathbb{N}_1$ (to be shown later), and it of course accepts some language $L^O_d$.

Suppose now $L^O_d$ were accepted by the $i$-th deterministic oracle Turing machine $D^O_i$ in the enumeration $e$, which is a deterministic $$T(n)=n^k+k$$ time-bounded oracle Turing machine. Let $D^O_i$ have $s$ states and $t$ tape symbols. Since $M^O_0$'s simulation can be done within time $$O(T(n)\log T(n)),\footnote{ See (III) of footnote \ref{footnote6}.}$$ we thus have that

\begin{align*}
\lim_{n\rightarrow\infty}&\frac{T(n)\log T(n)}{n^{k+1}}\\
=&\lim_{n\rightarrow\infty}\frac{(n^k+k)\log(n^k+k)}{n^{k+1}}\\
=&\lim_{n\rightarrow\infty}\Big(\frac{n^k\log(n^k+k)}{n^{k+1}}+\frac{k\log(n^k+k)}{n^{k+1}}\Big)\\
=&0\\
<&1.
\end{align*}

So, there exists an $N_0>0$ such that for any $N\geq N_0$, $$T(N)\log T(N)<N^{k+1},$$ which implies that for a sufficiently long $x$, say $|x|\geq N_0$, and $D^O_x$ denoted by such $x$ is $D^O_i$, we have that $$T(|x|)\log T(|x|)<|x|^{k+1}.$$
   
Thus, on input $x$, $M^O_0$ has sufficient time to simulate $D^O_x$ and accepts if and only if $D^O_x$ rejects. This is also because in the simulation of a polynomial-time deterministic oracle Turing machine with oracle $O$, $M^O_0$ only turns itself off mandatorily when the counter on tape $3$ overflows, i.e., when the counter $$\geq |x|^{k+1},$$ which happens after the end of the simulation. But, we assumed that $D^O_i$ accepted $L^O_d$, i.e., $D^O_i$ agreed with $M^O_0$ on all inputs. A contradiction.
      
We thus conclude from the above argument that $D^O_i$ does not exist in the enumeration $e$, i.e., $L^O_d$ is not accepted by any machine in the enumeration $e$. In other words, $$L^O_d\not\in \mathcal{P}^O.$$
      
By a similar argument appearing in the proof of Theorem \ref{theorem5}, it is not hard to see that for any input $w$ to $M^O_0$, there is a corresponding positive integer $i_w\in\mathbb{N}_1$ such that $M^O_0$ runs at most $|w|^{i_w+1}$ steps. Namely, $M^O_0$ runs within time $$O(n^k)$$ for all $k\in\mathbb{N}_1$.

Next we show that $L^O_d\in \mathcal{NP}^O$. Similarly, we define the family of languages \[\{L^O_{d,i}\}_{i\in\mathbb{N}_1}\] as follows:
\begin{align*}
         L^O_{d,i}\overset{{\rm def}}{=}&\text{ language accepted by $M^O_0$ running within time $O(n^i)$ for fixed $i\in\mathbb{N}_1$.}\\
         &\text{ That is, $M^O_0$ turns itself off mandatorily when its moves made by itself }\\
          &\text{ during the computation exceed $n^{i+1}$ steps.}
\end{align*}

\noindent Similarly, the above definition of language $L^O_{d,i}$ technically still can be done by adding a new tape to $M_0^O$ as a counter to count up to $$n^{i+1}$$ for a fixed $i\in\mathbb{N}_1$, meaning that $M_0^O$ turns itself off when the counter of tape $3$ exceeds $$n^{k+1}$$ or the counter of the newly added tape exceeds $$n^{i+1}.$$ 

Then by construction, i.e., since $M^O_0$ runs at most $|w|^{i_w+1}$ steps for any input $w$ where $i_w$ is a corresponding integer in $\mathbb{N}_1$, we thus have
\begin{equation}
     \label{eq7}
       \begin{split}
        L^O_d=&\bigcup_{i\in\mathbb{N}_1}L^O_{d,i}.
     \end{split}
\end{equation}

Furthermore, $$L^O_{d,i}\subseteq L^O_{d,i+1},\quad\text{for each fixed $i\in\mathbb{N}_1$}. $$ Since for any word $x\in L^O_{d,i}$ accepted by $M^O_0$ within $O(n^i)$ steps, it surely can be accepted by $M^O_0$ within $O(n^{i+1})$ steps, i.e., $$x\in L^O_{d,i+1}. $$

This gives that for any $i\in\mathbb{N}_1$,
\begin{equation}
       \label{eq8}
        \begin{split}
         L^O_{d,1}\subseteq L^O_{d,2}\subseteq\cdots\subseteq L^O_{d,i}\subseteq L^O_{d,i+1}\subseteq\cdots
       \end{split}
\end{equation}

Note further that for any fixed $i\in\mathbb{N}_1$, $L^O_{d,i}$ is the language accepted by a nondeterministic oracle Turing machine $M^O_0$ within $O(n^i)$ steps; we thus obtain 
\begin{equation}
     \label{eq9}
       \begin{split}
         L^O_{d,i}\in{\rm NTIME}^O[n^i]\subseteq\mathcal{NP}^O,\quad\text{for any fixed $i\in\mathbb{N}_1$}.
       \end{split}
\end{equation}

From (\ref{eq7}), (\ref{eq8}), and (\ref{eq9}), we deduce that $$L^O_d\in \mathcal{NP}^O.$$

To summarize, we obtain $$\mathcal{P}^O\neq\mathcal{NP}^O,$$ which contradicts the condition that $\mathcal{P}^O=\mathcal{NP}^O$. So, we can conclude that the set $P^O$ of all polynomial-time deterministic oracle Turing machines with oracle $O$ is not enumerable. This completes the proof of Theorem \ref{theorem2}. \Q.E.D

\vskip 0.3 cm
\begin{remark}
\label{remark6}
In fact, under the condition that $$\mathcal{P}^O=\mathcal{NP}^O,$$ we can suppose first that the set $P^O$ of all polynomial-time deterministic oracle Turing machines with oracle $O$ is enumerable. Then, we can show next that for any enumeration $$e:\mathbb{N}_1\rightarrow P^O,$$ there is always a machine $D^O_S$ that is in $P^O$, such that $$e(i)\neq D^O_S$$ for all $i\in\mathbb{N}_1$, thus contradicting the assumption that $P^O$ is enumerable.\footnote{ The language accepted by machine $D^O_S$ differs from the languages accepted by all of the polynomial-time deterministic oracle Turing machines in the enumeration, but it lies in $\mathcal{P}^O$ since $$\mathcal{P}^O=\mathcal{NP}^O.$$}
        
The reason why, for any enumeration of $P^O$ (when supposing $P^O$ is enumerable), there is always a machine $D^O_S$ in $P^O$ accepting the language $$L_d^O,$$ which is accepted by $M_0^O$ constructed in the proof of Theorem \ref{theorem2}, is that we are under the assumption that $\mathcal{P}^O=\mathcal{NP}^O$ and we have already shown the result $L_d^O\in\mathcal{NP}^O$ above, which leads to $$L_d^O\in\mathcal{P}^O.$$
        
The above arguments indicate that we are unable to diagonalize against the set of $P^O$ of all polynomial-time deterministic oracle Turing machines with oracle $O$, just as Cantor \cite{Can91} was unable to put all real numbers in the open interval $(0,1)$ into the slots indexed by all $i\in\mathbb{N}_1$.\footnote{ For a more detailed comparison, the reader could consult the second proof (due to Cantor) that the continuum is not enumerable. See \cite{Hob21}, p. 82.}
\end{remark}
\vskip 0.3 cm

By comparing the proof of Theorem \ref{theorem2} and Theorem \ref{theorem3}, we thus can conclude that the direct consequence of $$\mathcal{P}^O=\mathcal{NP}^O$$ does not qualify the technique of {\em diagonalization by a universal nondeterministic oracle Turing machine with oracle $O$} to separate $\mathcal{P}^O$ and $\mathcal{NP}^O$. In other words, {\em diagonalization techniques} ({\em via a universal nondeterministic oracle Turing machine}) would {\em not} apply to the relativized versions of the $\mathcal{P}$ versus $\mathcal{NP}$ problem because the set $P^O$ of all polynomial-time deterministic oracle Turing machines with oracle $O$ is not enumerable in this case, as can be seen from above that we can always construct a machine $D^O_S$ such that, for any function $$e:\mathbb{N}_1\rightarrow P^O,$$ there exists no $i\in\mathbb{N}_1$ such that $$e(i)=D^O_S.$$ Or equivalently, the cardinality of $P^O$ is greater than that of $\mathbb{N}_1$ (i.e., there is no $(1,1)$ correspondence between the set $P^O$ and $\mathbb{N}_1$), whereas the cardinality of the set $P$ of all polynomial-time deterministic Turing machines is equal to that of $\mathbb{N}_1$ in Section \ref{sec:diagonalization1}, which is the most significant difference. In brief, the fact that the set $P$ of all polynomial-time deterministic Turing machines being enumerable is an important prerequisite for the application of the {\em diagonalization techniques}.
      
Moreover, even if we suppose that $P^O$ is enumerable and $\mathcal{P}^O=\mathcal{NP}^O$, the {\em diagonalization techniques} ({\em via a universal nondeterministic oracle Turing machine}) would {\em not} apply as well, because {\em $M_0^O$ is also in $P^O$} in this case and $M_0^O$ is {\em unable to diagonalize against itself}.
      
\vskip 0.3 cm
\begin{remark}
As the reader can see, we follow all of Cantor's mathematical premises of the diagonalizing approach when proving Theorem \ref{theorem1} and Theorem \ref{theorem2}. And, as a matter of fact, following all of the mathematical premises is exactly what we need to do when applying the diagonalization techniques.
\end{remark}

\vskip 0.3 cm
\section{Concluding Remarks and Open Problems}
\label{sec:conclusions}
\vskip 0.3 cm

To summarize, we have shown that there exists a language $L_d$ that is accepted by some nondeterministic Turing machines but by no polynomial-time deterministic Turing machines. To achieve this, we first encode any single-tape deterministic Turing machine into a positive integer (i.e., a binary string) by using the method given in \cite{AHU74}. After that, the polynomial-time deterministic Turing machine $(M,k)$ could be the concatenation of the binary string representing $M$ itself followed by the order $10^k1$ of $(M,k)$. Our encoding of a polynomial-time deterministic Turing machine makes it very convenient for us to map any polynomial-time deterministic Turing machine to a positive integer, thus showing that the set $P$ of all polynomial-time deterministic Turing machines is enumerable, and simultaneously, the encoding makes it very convenient for us to control the running time, i.e., the time complexity of the simulating machine $M_0$ constructed in Section \ref{sec:diagonalization1}.

Next, we design a four-tape universal nondeterministic Turing machine $M_0$ that diagonalizes against all of the polynomial-time deterministic Turing machines. The Theorem \ref{theorem4} illustrates the operation of the universal nondeterministic Turing machine $M_0$ in detail, showing that there is a language $L_d$ accepted by this universal nondeterministic Turing machine $M_0$ but by no polynomial-time deterministic Turing machines. In Theorem \ref{theorem5}, we carefully analyze the running time of the universal nondeterministic Turing machine $M_0$, showing that it runs within time $$O(n^k)$$ for any $k\in\mathbb{N}_1$. We then demonstrate Theorem \ref{theorem6}, which states that $$L_d\in\mathcal{NP}. $$ Combining Theorem \ref{theorem4} and Theorem \ref{theorem6}, Theorem \ref{theorem1} hence follows. As we observed, the techniques applied in this work, as an interesting application, can be utilized to show that {\em one-way functions in the worst-case model} (see p. 281--284 in \cite{Pap94}, or see also \cite{Ko85,GS88}) do exist.

We have also presented a language $L_s$ that is in $\mathcal{P}$, and the machine accepting it also runs within time $$O(n^k)$$ for all $k\in\mathbb{N}_1$. We further conjecture that there is no deterministic $$O(n^c)$$ time-bounded Turing machine for fixed $c>0$ that is unable to accept $L_s$. Further, we have shown that under the conditions {\bf I} and {\bf II} given in Subsection \ref{subsec:main_results}, if $$\mathcal{P}^O=\mathcal{NP}^O,$$ then the set $P^O$ of all polynomial-time deterministic oracle Turing machines with oracle $O$ is not enumerable. So the reader can convince himself that we cannot use the {\em diagonalization techniques} by a universal nondeterministic oracle Turing machine with oracle $O$ to separate the complexity classes $\mathcal{P}^O$ and $\mathcal{NP}^O$ in the case of $$\mathcal{P}^O=\mathcal{NP}^O.$$ This shows that the so-called ``relativization barrier" is not really a barrier, but the polynomial-time oracle Turing machines, or the relativized versions of the $\mathcal{P}$ versus $\mathcal{NP}$ problem, are indeed somewhat mysterious.

There are many important questions that we did not touch on. For example, one question among them is the relationship between $\mathcal{NP}$ and ${\rm co}\mathcal{NP}$. Noting that we just mentioned in Section \ref{sec:diagonalization1} that we do not know whether $$\mathcal{NP}={\rm co}\mathcal{NP}.$$ Namely, are these two complexity classes the same? Also note that there is a subfield of computational complexity theory, namely, proportional proof complexity, which was initiated by Cook and Reckhow \cite{CR79} and is devoted to the goal of proving the conjecture $$\mathcal{NP}\neq {\rm co}\mathcal{NP}.$$ We refer the reader to the reference \cite{Coo00} for the importance of this research field and to the reference \cite{Kra95} for the motivation of the development of this rich theory. Apart from this, Chapter $10$ of \cite{Pap94} also contains the introductions of the importance of the problem $$\mathcal{NP}\overset{?}{=}{\rm co}\mathcal{NP}.$$ We also hope that the techniques developed in this work will shed some light on the proof of this important conjecture.

Another intriguing open question is whether the cardinality of the set $P^O$ of all polynomial-time deterministic oracle Turing machines with oracle $O$ is less than that of real numbers if \[\mathcal{P}^O=\mathcal{NP}^O?\,\footnote{ It is a question about the {\it Continuum hypothesis}. See \cite{A3}. Note that we have already demonstrated that the cardinality of the set $P^O$ of all polynomial-time deterministic oracle Turing machines with oracle $O$ (when $\mathcal{P}^O=\mathcal{NP}^O$) is greater than that of positive integers, under the assumptions (I), (II), and (III) in {\it footnote} \ref{footnote6} (i.e., if the assumptions (I), (II), and (III) in {\it footnote} \ref{footnote6} are satisfied). So to attack such an important and interesting question, one should rigorously show that the assumptions (I), (II), and (III) are valid. }\]

Finally, although today's or future's computers (which are similar to that described in {\it footnote} \ref{footnote28}), which fall into the category of polynomial-time universal deterministic Turing machines (not the polynomial-time universal probabilistic Turing machines nor the polynomial-time universal quantum Turing machines), cannot accurately calculate the entire $\mathcal{NP}$ set of problems, do not be pessimistic; our computers can solve as many problems as the cardinality of $\mathbb{N}_1$. Looking forward to discovering more and more practical polynomial-time deterministic algorithms to make our living environment much more convenient and much more colorful.

\section*{Acknowledgements}
\label{sec:acks}
Sincere thanks go to complexity theorist Lance Fortnow \cite{For21} for his valuable criticisms, which cause us to seek Theorem \ref{theorem6}.

\bibliographystyle{aomplain}

\appendix
\vskip 0.3 cm
\section{The Two Definitions of $\mathcal{P}$ (Respectively, $\mathcal{NP}$) Are Equivalent}
\label{sec:appendix}
\vskip 0.3 cm

The official descriptions of the definitions of $\mathcal{P}$ and $\mathcal{NP}$ in \cite{Coo00} are given by
 \begin{definition}
\label{defP}
   $$
      \begin{array}{ll}
        \mathcal{P}\overset{{\rm def}}{=}&\{L\,|\,L=L(M)\text{ for some Turing machine $M$ that runs} \\
         &\qquad\text{ in polynomial time }\}.
      \end{array}
   $$
\end{definition}
   
In Definition \ref{defP}, Turing machine $M$ refers to a deterministic Turing machine, and the notion of ``running in polynomial time" is that if there exists $k$ ($\in\mathbb{N}_1$) such that, $$T_M(n)\leq n^k+k,\quad\forall n\in\mathbb{N}_1$$ where $T_M(n)$ is defined by $$T_M(n)=\max\{t_M(w)\,|\,w\in\Sigma^n\}$$ and $t_M(w)$ denotes the number of steps (or moves) in the computation of $M$ on input $w$ of length $n$.
\begin{definition}\label{defNP}
$$\mathcal{NP}\overset{{\rm def}}{=}\{L\,|\,w\in L\Leftrightarrow\exists y(|y|\leq|w|^k\,\,\,{\rm and }\,\, R(w,y))\} $$ where the {\it checking relation} $$R(w,y)$$ is polynomial-time, i.e., the language $$L_R\overset{{\rm def}}{=}\{w\#y\,|\,R(w,y)\}$$ is in $\mathcal{P}$.
\end{definition}

As we mentioned earlier, these two definitions look a little different from those given in Section \ref{sec:preliminaries}. But they are equivalent. Here, we will show the equivalence for these two definitions of $\mathcal{P}$ and the equivalence for those two definitions of $\mathcal{NP}$.
  
\vskip 0.3 cm
\subsection{A Proof of Two Definitions of $\mathcal{P}$ Are Equivalent}
\label{subsec:proof_p_equivalent}
\vskip 0.3 cm
  
Now, we are at an opportune point to show that the above two definitions of $\mathcal{P}$ are equivalent.
  
\vskip 0.3 cm
\begin{proof}
We show first the ``if" part. Suppose that $$L\in\bigcup_{i\in\mathbb{N}_1}{\rm DTIME}[n^i]. $$ Then there is a $k\in\mathbb{N}_1$, such that $$L\in{\rm DTIME}[n^k] $$ which means that for all $n\in\mathbb{N}_1$, there is a deterministic Turing machine $M$, for any $w\in\Sigma^n$ $$T_M(|w|)\leq c_0n^k+c_1n^{k-1}+\cdots +c_{k-1}n+c_k\quad\text{where $c_0>0$},$$ and $$L=L(M). $$ For such constants $c_0$, $c_1$, $\cdots$, $c_k$, there must exist a minimal $t\in\mathbb{N}_1$ such that for all $n\in\mathbb{N}$ and for any $w\in\Sigma^n$ $$T_M(|w|)\leq n^t+t. $$ So, $$L\in\{L\,|\,L=L(M)\text{ for some Turing machine $M$ that runs in polynomial time}\}. $$
  
\vskip 0.3 cm
We show next the ``only if" part. Suppose now that the language $$L\in \{L\,|\,L=L(M)\,\text{ for some Turing machine $M$ that runs in polynomial time}\}.$$ Then there exists a $k\in\mathbb{N}_1$ such that for all $n\in\mathbb{N}_1$ and for all $w\in\Sigma^n$, $$T_M(|w|)\leq n^k+k, $$ which implies that $$L\in{\rm DTIME}[n^k]\subseteq\bigcup_{i\in\mathbb{N}_1}{\rm DTIME}[n^i]. $$ The conclusion follows.
\end{proof}
  
\vskip 0.3 cm
\subsection{A Proof of Two Definitions of $\mathcal{NP}$ Are Equivalent}
\vskip 0.3 cm
  
The remainder of this appendix is to show that {\em Definition} \ref{defNP} and the definition of $\mathcal{NP}$ given in Section \ref{sec:preliminaries} are equivalent.

To begin, we should first give another definition of $\mathcal{NP}$, which is similar to the official descriptions of the definition of $\mathcal{P}$ in \cite{Coo00} (i.e., Definition \ref{defP} above):
\begin{definition}   
\label{defaNP}
$$
  \begin{array}{ll}
     \mathcal{NP}\overset{{\rm def}}{=}&\{L\,|\,L=L(M)\text{ for some nondeterministic Turing machine $M$} \\
     &\qquad\text{that runs in polynomial time}\}.
  \end{array}
$$
\end{definition}
   
Now, let us finish the last step of the task that we mentioned at the beginning of this subsection.
\vskip 0.2 cm
\begin{proof}
We need to show that Definition \ref{defaNP} is equivalent to Definition \ref{defNP} and to show the equivalence between Definition \ref{defaNP} and the definition of $\mathcal{NP}$ given in Section \ref{sec:preliminaries}. By a similar argument given in Subsection \ref{subsec:proof_p_equivalent}, it is clear that Definition \ref{defaNP} and the definition of $\mathcal{NP}$ given in Section \ref{sec:preliminaries} are equivalent; thus, this part of the proof will be omitted for brevity.
    
It is time for us to show that Definition \ref{defaNP} is equivalent to Definition \ref{defNP}. For the purposes of brevity, we omit the proof and refer the reader to \cite{Kar72} (cf. Theorem 1 in \cite{Kar72}), or to the proof of Theorem $7$.$20$ in \cite{Sip13}, which is a theorem answering the same question discussed here (see p. $294$ of \cite{Sip13}). Thus, this completes the proof.
\end{proof}
\vskip 0.5 cm
\end{document}